\documentclass[fleqn,usenatbib]{mnras}

\usepackage{newtxtext,newtxmath}

\usepackage[T1]{fontenc}

\DeclareRobustCommand{\VAN}[3]{#2}
\let\VANthebibliography\thebibliography
\def\thebibliography{\DeclareRobustCommand{\VAN}[3]{##3}\VANthebibliography}

\usepackage{graphicx}	
\usepackage{amsmath}	
\usepackage{macros_vdbosch}
\usepackage{orcidlink}

\graphicspath{{./}{figures/}}

\title[SIDM promotes bar formation]{Self-interacting dark matter promotes bar formation in disk galaxies}

\author[S. Dattathri et al.]{
Shashank Dattathri$^{1}$\thanks{E-mail: shashank.dattathri@yale.edu}\orcidlink{0000-0002-7941-1149},
Frank C. van den Bosch$^{1}$\orcidlink{0000-0003-3236-2068},
HanYuan Zhang$^{2}$\orcidlink{0009-0005-6898-0927},
Martin D. Weinberg$^{3}$\orcidlink{0000-0003-2660-2889},
\newauthor
Eugene Vasiliev$^{4}$\orcidlink{0000-0002-5038-9267},
Priyamvada Natarajan$^{1}$\orcidlink{0000-0002-5554-8896},
Vasily Belokurov$^{3}$\orcidlink{0000-0002-0038-9584}\\\\
$^1$Department of Astronomy, Yale University, PO. Box 208101, New Haven, CT 06520-8101\\
$^2$Institute of Astronomy, University of Cambridge, Madingley Road, Cambridge CB3 0HA, UK\\
$^3$Department of Astronomy, University of Massachusetts, Amherst, MA01003-9305, USA\\
$^4$University of Surrey, Guildford, GU2 7XH, UK
}

\date{Accepted XXX. Received YYY; in original form ZZZ}

\pubyear{\the\year{}}

\begin{document}
\label{firstpage}
\pagerange{\pageref{firstpage}--\pageref{lastpage}}
\maketitle

\begin{abstract}
Despite its remarkable success on large scales, the standard $\Lambda$CDM paradigm faces persistent small-scale challenges that have motivated alternative models for the dark sector. Self-interacting dark matter (SIDM) offers a compelling possibility, in which dark matter particles can scatter off each other. Stellar bars are a ubiquitous feature of disk galaxies across cosmic time. Bars are dynamically coupled to their host galaxy’s dark matter halo, and therefore their properties provide a powerful probe of the nature and distribution of dark matter. In this paper, we use idealized, high-resolution $N$-body simulations and analytic calculations based on kinetic theory to study bar formation and evolution in disk galaxies embedded in SIDM halos. We find that compared with collisionless CDM, SIDM produces bars that form earlier and grow to larger amplitudes, even for modest self-interaction cross sections. In several cases, disks that remain stable in CDM, including kinematically hot and dark-matter-dominated disks, develop strong bars in SIDM. This accelerated bar growth occurs because self-interactions broaden the bar-halo resonances and enhance angular momentum transfer from the stellar disk to the halo. We explicitly show that this phenomenon is not related to core formation in SIDM halos. At late times, gravothermal core collapse can raise the central dark matter density enough to weaken or dissolve the bar. These results suggest that the abundance, strength, and redshift evolution of barred galaxies offer a promising observational route to constraining dark matter self-interactions, particularly in light of the growing sample of high-redshift bars revealed by JWST.
\end{abstract}




\section{Introduction}
\label{sec:intro}

Stellar bars found in 50-65\% of disk galaxies in the local universe \citep{Erwin2018}, making them a ubiquitous feature. Bars are important drivers of galaxy evolution. Among other things, they redistribute angular momentum within the galaxy, cause radial migration, induce star formation, and result in the formation of rings and pseudobulges \citep[see][for reviews]{Kormendy2004, Sellwood2014}. In addition, bars provide one of the most efficient internal mechanisms for transporting gas inward toward the central few hundred parsecs, where it can lead to nuclear star formation and potentially fuel the central supermassive black hole \citep{Shlosman1989,Jogee2006, HopkinsQuataert2010}.

The formation and evolution of bars has been a widely researched topic for several decades. In a seminal paper, \citet{Ostriker1973} showed that flat disk galaxies, which are violently unstable to bar formation, can be stabilized by embedding them inside a larger spherical halo. They concluded that the fact that not all disk galaxies host bars provides strong evidence for the presence of dark matter. Later, \citet{Efstathiou1982} used a series of N-body experiments to derive an empirical relation describing bar instability, a relation that is widely used in semi-analytic models \citep[e.g.][]{Mo1998,vandenbosch1998,Benson2010}. 

However, it is important to note that the aforementioned papers modeled the dark matter halo as a static potential. In reality, bars and halos are dynamically coupled, and hence the halo's response to the rotating bar can contribute significantly to its evolution. The analytical theory for bar-halo interactions was established in two seminal papers by \citet{LBK} and \citet{TW84}. This prompted a number of studies on this topic using numerical simulations \citep{Debattista1998,Debattista2000,HolleyBockelmann2005, Petersen2016}, analytical calculations \citep{Weinberg1985, Chiba2022, Chiba2023, Hamilton2023}, as well as combination and comparison between the two \citep{Hernquist1992, Weinberg2007,Weinberg2007b}. In particular, \citet{Athanassoula2002} showed that compared to static halos, live halos are able to respond to bar instabilities and promote bar formation, highlighting the importance of bar-halo coupling. 

It has since become clear that in the real universe, bar formation and evolution is a complex process that depends on many factors. The various properties of the dark matter halo, including triaxiality \citep{Machado2010}, the inner density slope \citep{Berentzen2006}, spin \citep{Saha2013,Kataria2025}, and stellar-to-halo mass ratio \citep{Reddish2022,Lopez2024} all influence the presence and properties of bars. In addition, the presence of gas can affect angular momentum transfer between the disk and halo, leading to delayed bar growth (\citealt{Athanassoula2013}, but see also \citealt{BlandHawthorn2024}) and can prevent bar slowdown at late times \citep{Friedli1993,Beane2023}. Bars can also form through alternative mechanisms, such as tidal interactions \citep{Noguchi1987,Gerin1990,Mayer2001, Lokas2014, Merrow2024, Kodama2026} and minor mergers \citep{Kazantzidis2008}. Finally, central mass concentrations like supermassive black holes can also play an important role, leading to bar dissolution \citep{Shen2004, Du2017} or strengthening \citep{Wheeler2023}. Given the complex interplay between these various factors, it is no surprise that there is little convergence in bar properties or their population statistics in different cosmological simulations \citep{Zhou2020, Roshan2021, Reddish2022}. 

JWST has been revolutionizing our understanding of galaxy formation and evolution, particularly in the high redshift regime. One of its striking findings is the discovery of barred galaxies at high redshift, as early as $z \sim 4$ \citep{Guo2023,Costantin2023,LeConte2024,Guo2025, Geron2025,LeConte2026, Ivanov2026}. These discoveries have challenged the conventional idea that bar emergence is a low redshift phenomenon requiring well-settled dynamically cold disks. While it is unclear whether high redshift bars, and other JWST observations such as the early emergence of supermassive black holes \citep{Maiolino2024}, are necessarily in tension with the standard $\Lambda$CDM paradigm, they have motivated exploration of alternative models. 

Self-interacting dark matter is a particularly intriguing alternative dark matter model that predicts that dark matter particles interact, or scatter, with each other \citep{Spergel2000}. SIDM was originally proposed as a solution to several small-scale tensions with CDM \citep[see][for a review]{Bullock2017}. Recent studies have invoked SIDM to explain the diversity of dwarf galaxy rotation curves \citep{Kamada2017,Roberts2025,Zeng2025,Gutcke2025}, the galaxy-galaxy strong lensing discrepancy \citep{Dutra2025, Natarajan2026}, and the early seeding of supermassive black holes \citep{Pollack2015,Jiang2026}.

In this paper, we use idealized high-resolution N-body simulations complemented by analytic calculations based on kinetic theory to study the formation and evolution of barred galaxies in SIDM halos. We show that disk galaxies embedded in SIDM halos are able to form strong bars significantly faster than in CDM halos. We also demonstrate cases in which some disks that are stable against bar formation in CDM for over a Hubble time form a strong bar in SIDM within a few Gyr. This is because dark matter self-interactions accelerate angular momentum transfer from the disk to the halo by broadening the bar-halo resonances. These results offer a possible explanation for the high bar fraction observed in the early universe by JWST. We also argue that future surveys of bar population statistics across cosmic time have the potential to constrain the strength of dark matter self-interactions. Given their role in driving gas toward galactic centers, the enhanced bar formation we find in SIDM halos also has important implications for the buildup of nuclear bulges and  fueling history of supermassive black holes across cosmic time.

This paper is organized as follows. In Section~\ref{sec:methodology}, we describe our initial conditions and the methodology used for the simulations. Section~\ref{sec:barstrength} presents the evolution of the bar strength over time in the various cases. The underlying dynamics at play, and how they differ between CDM and SIDM, are studied in Section~\ref{sec:resonances}. We discuss the implications of our results and conclude in Section~\ref{sec:conclusion}.

\section{Methodology}
\label{sec:methodology}

\subsection{Initial Conditions}

The initial conditions for all the models discussed in this paper are a composite disk-bulge-halo system, constructed using the self-consistent modeling method implemented in the \texttt{Agama} software \citep{Vasiliev2019}\footnote{\url{https://github.com/GalacticDynamics-Oxford/Agama/blob/master/py/example_self_consistent_model_bdh.py}}, as described in section~5 of that paper.

The spherical dark matter halo is represented by a generalized NFW profile:
\begin{equation}
\label{eq:nfw}
    \rho(r) = \frac{\rho_h}{\left(\frac{r}{r_s}\right)^\gamma\left(1+\frac{r}{r_s}\right)^{3-\gamma}} \times \exp{\left(-\frac{r}{r_{\rm cut}} \right)}
\end{equation}
where the second term represents a smooth truncation at large radii, and $\gamma$ controls the inner density slope of the halo. We explore $\gamma=1$, which corresponds to the NFW profile \citep{Navarro1997}, and $\gamma=0$, which yields a central constant density core. Throughout, the scale radius $r_s$ is held fixed at $20$ kpc, the cutoff radius $r_{\rm cut}$ is set to 200 kpc, and the normalization factor $\rho_h$ is adjusted such that the total mass of the halo equals $10^{12} M_\odot$, which crudely correspond to the values for the Milky Way halo \citep{McMillan2017}.

The disk is modeled as a radial exponential profile with an isothermal vertical profile:
\begin{equation}
    \rho(R,z) = \Sigma_0 \exp{\left(-\frac{R}{R_d} \right)} \times \frac{1}{4h} \sech^2\left(\frac{z}{2h} \right) \ ,
\end{equation}
where $\Sigma_0$ is the normalization parameter (which sets the disk mass), $R_d$ is the scale radius of the disk, and $h$ is the vertical scale height. Throughout, we fix $R_d=2.5$ kpc and $h=0.3$ kpc, which again roughly correspond to the Milky Way values. 

The central bulge is represented by a simple truncated power-law:
\begin{equation}
    \rho(r) = \rho_b \left(\frac{r}{r_b} \right)^{-1} \times \exp \left(-\frac{r}{r_{\rm cut}} \right) \ ,
\end{equation}
with $r_b=r_{\rm cut}=1$ kpc.

\begin{table*}
\caption{List of simulations analyzed in this paper. Across runs, all the properties of the dark matter halo except its inner density slope $\gamma$ are held fixed. The radial and vertical scale lengths of the stellar components are also constant. The parameters {\tt f\_stars} and {\tt f\_bulge} are the fractional contribution of the stars and the bulge respectively to the total rotation curve, which indirectly set the stellar and bulge mass of the galaxy. The velocity dispersion profile of the disk is determined by $Q_{\rm min}$, the minimum value of the Toomre $Q$ parameter. Each model suite consists of three simulations: CDM, $\sigma_\rmm=1\cmg$, and $\sigma_\rmm=10\cmg$ (the {\tt Fiducial} suite has an extra run with $\sigma_\rmm=0.1\cmg$). The bar formation timescale $T_{\rm bar}$ refers to the time at which the bar amplitude $A_2/A_0$ reaches $0.15$, at which point the bar is qualitatively visible in the surface brightness maps.}
\centering
\label{tab:sims}
\begin{tabular}{|c|c|c|c|c|c|c|c|c|p{5cm}|}
\hline
Name & 
\begin{tabular}{c}
$M_{\rm stellar}$ \\
$(10^{10}\,M_\odot)$
\end{tabular} &
\texttt{f\_stars} &
\texttt{f\_bulge} &
$Q_{\rm min}$ &
$\gamma$ &
\multicolumn{3}{c|}{$T_{\rm bar}$ (Gyr)} & Notes \\
\cline{7-9}
 &  &  &  &  &  & CDM & $1 \cmg$ & $10\cmg$ & \\
\hline

{\tt Fiducial} & 3.88 & 0.5 & 0 & 2.5 &1.0 & 11.08 & 6.28 & 3.20 & Fiducial high-resolution simulation \\ \hline

{\tt Model A} & 0.99 & 0.2 & 0.2 & 1.0 &1.0 & -- & -- & 6.95 & Dark matter dominated system, does not form a bar in CDM or $\sigma_\rmm=1\cmg$\\
\hline

{\tt Model B} & 1.67 & 0.3 & 0.1 & 1.0 &1.0 & 11.76 & 10.60 & 3.08 & \\
\hline

{\tt Model C} & 1.69 & 0.3 & 0.2 & 1.0 &1.0 & 10.00 & 8.12 & 3.16 & \\
\hline

{\tt Model D} & 2.61 & 0.4 & 0.1 & 2.5 &1.0 & 10.20 & 5.96 & 2.72 & \\
\hline

{\tt Model E} & 2.61 & 0.4 & 0.2 & 3.0 &1.0 & -- & 11.04 & 3.00 & Kinematically hot disk, does not form a bar in CDM \\
\hline

{\tt Model F} & 5.80 & 0.6 & 0.1 & 3.0 &1.0 & 11.36 & 7.80 & 3.76 & Close to present-day Milky Way stellar mass\\ 
\hline 

{\tt Core} & 1.58 & 0.6 & 0 & 2.0 &0.0  & -- & 5.76 & 5.84 & Initially cored dark matter halo\\ 
\hline

{\tt Freeze} & 3.88 & 0.5 & 0 & 2.5 &1.0 & \multicolumn{3}{c|}{--} & {\tt Fiducial}'s $\sigma_\rmm=10 \cmg$ run with self-interactions turned off at $T=0.4$ and 0.8 Gyr (see Section~\ref{ssec:verify})\\ 
\hline
\end{tabular}
\end{table*}

We set the density normalizations of the disk and bulge as follows, motivated by e.g. \citet[][]{Fujii2018,TepperGarcia2021}. The total circular rotation velocity at radius $8$ kpc is fixed at $v_{0}=240$ km/s. We then set a parameter {\tt f\_stars} which determines the contribution of the stars (disk+bulge) to $v_{0}$: 
\begin{equation}
    {\tt f\_stars} = \frac{ \left.\big({\partial \Phi_{\rm stars}}\big/{\partial R}\big)\right|_{8 \ {\rm kpc}} }{\left.\big({\partial \Phi_{\rm tot}}\big/{\partial R}\big)\right|_{8 \ {\rm kpc}}} \ ,
\end{equation}
where $\Phi_{\rm stars}$ and $\Phi_{\rm tot}$ are the stellar and total gravitational potentials, respectively. This determines the total stellar mass of the galaxy, which is then further divided into the disk and bulge components using the parameter {\tt f\_bulge}${}=M_{\rm bulge}/M_{\rm stars}$.

The parameters {\tt f\_stars} and {\tt f\_bulge} are varied across runs (see Table~\ref{tab:sims}), yielding different stellar masses for the galaxy while keeping the total DM mass fixed. 

Finally, a key parameter that determines the bar formation timescale is the Toomre $Q$ parameter, defined as:
\begin{equation}
    Q(R) = \frac{\sigma(R)  \kappa(R)}{3.36 G \Sigma(R)} \ ,
\end{equation}
\citep{Toomre1964}, where $\sigma(R)$ is the radial velocity dispersion, $\kappa(R)$ is the radial epicyclic frequency, and $\Sigma(R)$ is the surface density of the stellar disk. We set the initial velocity profile of the disk such that the radial velocity  dispersion decreases exponentially as 
\begin{equation}
    \sigma(R)=\sigma_0 \exp{(-R/{\rm 6 \ kpc})} \ .
\end{equation}

The value of $\sigma_0$ is determined by setting the minimum value of the Toomre $Q$ parameter across the stellar disk to $Q_{\rm min}$, a free parameter. We explore values in the range $1 \leq Q_{\rm min} \leq 3$, which are conducive to bar formation on a timescale of $\sim$few Gyr in CDM \citep{Chen2025}.

Note that in setting up the initial conditions, the halo's density profile is fixed to that of Equation~(\ref{eq:nfw}), while its velocity distribution is computed using the combined halo+disk+bulge potential. Since the stars dominate the central potential, the central velocity dispersion of the halo is higher than that of the isolated NFW case. The velocity dispersion profile is the key determinant of the rate and direction of conductive heat transfer in SIDM. Therefore, all halos in this study reach core collapse earlier than in the corresponding dark matter only runs, in agreement with previous studies that demonstrate that the presence of baryons accelerates the gravothermal evolution of SIDM halos \citep[e.g.][]{Elbert2018, Zhong2023, vdb2025}.

\subsection{Implementation of self-interactions}

To implement the self-interactions between dark matter particles, we closely follow the method of \citet{vdb2025}, which is based on the models of \citet{Kochanek2000} and \citet{Vogelsberger2012}. We adopt a constant SIDM cross section per unit mass, $\sigma_\rmm$, and the self-interactions are restricted to isotropic elastic scattering. Only the DM particles undergo scattering, while the stellar particles remain collisionless. 

At each timestep $\Delta t$, the number of expected scattering events for each (DM) particle is computed as:
\begin{equation}
    \langle N_{\rm scat} \rangle = \frac{1}{2V_i} \sum_j \sigma_\rmm m_p \left| \mathbf{v}_i - \mathbf{v}_j \right| \Delta t \ , 
\end{equation}
where the summation $j$ runs over the $k$-nearest neighbors of particle $i$, $\mathbf{v}$ is the particle velocity, $m_p$ is the particle mass, and $V_i$ is the volume of the sphere centered on particle $i$ enclosing its $k$-nearest neighbors. This is equivalent to using the spherical top-hat kernel in \citet{vdb2025}. Throughout we set $k=32$, which is a value commonly used in SIDM simulations \citep[e.g.][]{Mace2024, Palubski2024}.

Once $\langle N_{\rm scat} \rangle$ is calculated, the actual number of scattering events $N_{\rm scat}$ for particle $i$ is determined by drawing from a Poisson distribution with an expectation value $\langle N_{\rm scat} \rangle$. If a particle undergoes scattering (i.e. $N_{\rm scat} \geq 1$), then for each scattering event, it is scattered off one of its 32 nearest neighbors, where the probability to scatter with particle $j$ is:
\begin{equation}
    P_{ij} = \frac{\left| \mathbf{v}_i-\mathbf{v}_j \right|}{\sum_j \left| \mathbf{v}_i-\mathbf{v}_j \right| } \ .
\end{equation}

After a particle $j$ has been chosen, the particles $i$ and $j$ are assumed to have undergone an elastic collision, and their velocities are updated as
\begin{equation}
\bv_i = \bv_{\rm cm} + (\bv_{ij}/2) \, \hat{\be} \, \nonumber 
\end{equation}
\vskip -0.55truecm
\begin{equation}
\bv_j = \bv_{\rm cm} - (\bv_{ij}/2) \, \hat{\be} \,,
\end{equation}
where $\bv_{ij} = \bv_j - \bv_i$, $\bv_{\rm cm}$ is the center of mass velocity of the pair, and $\hat{\be}$ is a random unit vector drawn from a sphere. We refer the reader to \citet{vdb2025} for more details and validation tests that demonstrate that our treatment of self-interactions is adequate.

\subsection{N-body setup}

Our CDM simulations are run using the publicly available code {\tt GyrfalcON} \citep{Dehnen2002}. All particles are softened with a gravitational softening length of $100$ pc ($50$ pc for the high-resolution {\tt Fiducial} suite). The simulations are run using hierarchical timestepping, such that each particle's timestep is set by its instantaneous acceleration $a_i$ and softening length as $\Delta t_i \sim 0.1 \sqrt{\epsilon/\left| a_i \right|}$. 

The SIDM simulations are run using a modified version of {\tt pyfalcon}\footnote{\url{https://github.com/GalacticDynamics-Oxford/pyfalcon}}, a Python interface for the same gravity solver {\tt falcON}, first introduced in \cite{Vasiliev2022}. At each timestep, the self-interactions are implemented as described in the previous subsection. The important difference is that in the SIDM runs, all particles are evolved on the same global timestep. While this results in a significant increase in computation time, we find that this ensures proper convergence in the core collapse time across simulations. The timestep itself is given by the minimum of the ``gravitational timestep'' and ``self-interactions timestep'':
\begin{equation}
    \Delta t={\rm min}(\Delta t_{\rm grav},\Delta t_{\rm SI}) \ , 
\end{equation}
where
\begin{equation}
    \Delta t_{\rm grav}={\rm min} \left( 0.1 \sqrt{\epsilon/\left| a_{\rm max} \right|}\right) \ , 
\end{equation}
where $\left| a_{\rm max} \right|$ is the largest magnitude of acceleration among all particles, and 
\begin{equation}
    \Delta t_{\rm SI}=\frac{P_{\rm max}}{{\rm max} \left( \frac{1}{2V_i} \sum_j \sigma_\rmm m_p \left| \mathbf{v}_i - \mathbf{v}_j \right|\right)} \ .
\end{equation}
The values of $\Delta t_{\rm grav}$ and $\Delta t_{\rm SI}$ are recomputed at every global timestep from the particles' instantaneous accelerations and 32 nearest neighbors. Here, $P_{\rm max}$ is the maximum expected number of scatterings per particle per timestep, which we set to $0.5$. It is noteworthy that this is significantly higher than the value adopted in other SIDM simulations \citep[e.g.][]{Vogelsberger2019,Palubski2024,Mace2024}. For example, \citet{Palubski2024} find that a value of $ P_{\rm max} \lesssim 0.02$ is necessary for sufficient energy conservation. However, since we do not use hierarchical time-stepping and run all simulations on a single compute core, the energy conservation in our simulations is $\left| \Delta E \right|/E \lesssim 10^{-3}$ even with our adopted $P_{\rm max}=0.5$, up to the late stages of core collapse. A direct one-to-one comparison between these studies is beyond the scope of this paper (but see the discussion in \citealt{vdb2025}). 

\begin{figure*}
    \centering
    \includegraphics[width=\textwidth]{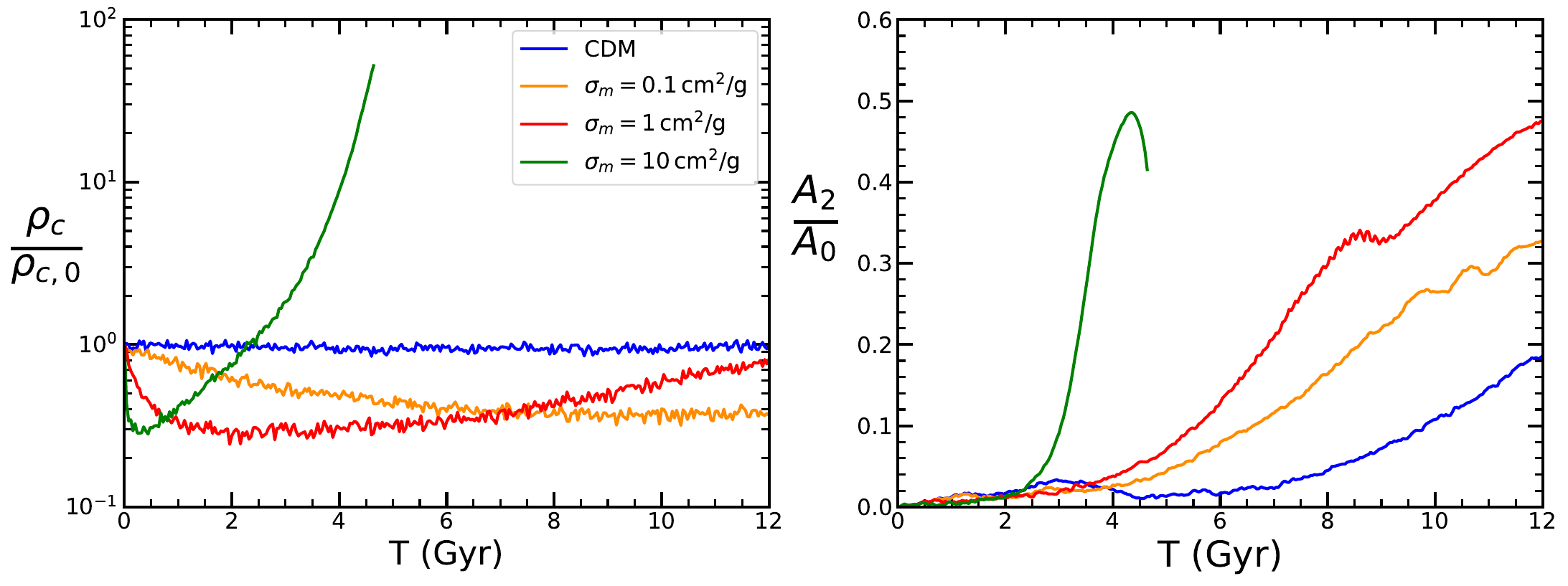}
    \caption{ {\it Left}: evolution of the central dark matter density ($\rho_c$) over time for the {\tt Fiducial} suite of simulations, normalized by its initial value ($\rho_{c,0}$). The central density remains constant in the CDM run, while in the SIDM runs, the central density initially drops (core formation) and then rises over time (core collapse). {\it Right}: bar amplitude ($A_2/A_0$) {\it vs} time for the same set of runs. In CDM, the disk is stable against bar formation until $\sim 7$ Gyr, and forms a moderately strong bar by the end of the simulation. In the SIDM runs, the bar forms earlier and reaches a higher amplitude, and this is more pronounced with increasing SIDM cross section. In the case of $\sigma_\rmm=10\cmg$, the amplitude has started to decrease at the end, before the simulation is terminated due to core collapse.}
    \label{fig:fiducial}
\end{figure*}

\subsection{List of simulations}

Table \ref{tab:sims} lists the various simulations discussed in the paper. The main parameters changed across runs are the total stellar mass and its split between the disk and the bulge, the minimum Toomre $Q$ parameter $Q_{\rm min}$, and the inner slope of the DM halo $\gamma$. We also list the bar formation time $T_{\rm bar}$. This is defined as the time when the bar amplitude (defined below) reaches 0.15, roughly corresponding to when the bar is qualitatively visible in the projected surface brightness maps. Across models, all of the halo parameters are held fixed, as well as the scale lengths for the stellar components ($R_d$, $h$, and $r_b$). We caution that our parameter space exploration is far from exhaustive. Some of the key factors affecting bar formation and evolution that we do not explore in this study are the presence of gas \citep{Berentzen2007,Villa-Vargas2010,BlandHawthorn2023} and the associated feedback \citep{Zana2019,Irodotou2022,Weinberg2025}, dark matter halo spin \citep{Kataria2025, Chen2025b}, and spatial extent of the disk \citep{Efstathiou1982, Klypin2009, Chen2025}.

Our {\tt Fiducial} simulations have $5\times10^6$ dark matter particles and $2.5\times 10^6$ star particles, which has been shown to yield convergence in both the SIDM halo evolution \citep{Mace2024} and disk/bar evolution \citep{Fujii2011, Zhou2026, Kwak2026}. All other simulations are run with $10^6$ dark matter particles and $5 \times 10^5$ star particles. We conducted tests by running simulations across two orders of magnitude in resolutions, and find that the overall conclusion of SIDM accelerating bar growth is robust, although the exact $A_2/A_0$ vs $T$ curve showed variation of around $\pm 1$ Gyr across runs (see \citealt{Sellwood2009}).

For each model, we run a set of three simulations: one CDM run where the DM particles are collisionless, and two SIDM runs with cross sections $\sigma_\rmm=1 \cmg$ and $\sigma_\rmm=10 \cmg$. The {\tt Fiducial} suite contains an additional run with $\sigma_\rmm=0.1 \cmg$. If possible, all the simulations are evolved until $12$ Gyr. However in SIDM, once the halo reaches core collapse, the simulation timesteps become extremely small due to the high density and accelerations in the core. Therefore the SIDM simulations are terminated when the central density has reached $\sim 30$ times its initial value.
This takes longer than a Hubble time for $\sigma_\rmm=0.1 \cmg$ and $1 \cmg$, but typically occurs within $\sim$ few Gyr for $\sigma_\rmm=10 \cmg$.

\section{Bar growth in CDM vs SIDM}
\label{sec:barstrength}

We define the bar amplitude in our simulations as the $m=2$ Fourier amplitude of the stellar disk normalized by its $m=0$ amplitude, i.e.:
\begin{equation}
    \frac{A_{2}}{A_0}=\left| \frac{\sum_i m_i \exp{(2i\phi_i)}}{\sum_i m_i} \right| \ , 
\end{equation}
where $m_i$ and $\phi_i$ are the mass and azimuthal angle of each particle respectively, and the summation runs over all star particles. 

\subsection{ {\tt Fiducial} simulation}

We begin with analyzing our {\tt Fiducial} set of simulations. The left panel of Figure \ref{fig:fiducial} shows the evolution of the central DM density $\rho_c$ (the average density within 0.5 kpc) over time, normalized by its initial value $\rho_{c,0}$, for CDM (blue), $\sigma_\rmm=0.1 \cmg$ (orange), $\sigma_\rmm=1\cmg$ (red) and $\sigma_\rmm=10 \ {\rm cm^2/g}$ (green). In the CDM run, the central density remains mostly constant over time, reflecting the stability of the NFW cusp. For $\sigma_\rmm=0.1\cmg$, the central density decreases over time, marking the core formation phase of SIDM evolution, characterized by inward heat transfer and erosion of the $\rho \propto r^{-1}$ NFW cusp. For $\sigma_\rmm=1 \cmg$, the central density initially decreases over time due to core formation, after which it starts to increase with time. This is due to outward heat transfer, resulting in the onset of gravothermal core collapse. The central density of the $\sigma_\rmm=10\cmg$ run shows similar behavior, except that the evolution is significantly faster. The simulation is terminated at $T \approx 4.6$ Gyr, at which point it has reached a late stage of core collapse.

The right panel of Figure \ref{fig:fiducial} shows the bar amplitude $A_2/A_0$ versus time for the {\tt Fiducial} set of runs. In the CDM run, the bar amplitude remains close to zero for $T \approx 6$ Gyr, after which it continuously rises until the end of the simulation. In the $\sigma_\rmm=0.1\cmg$ run, bar growth is accelerated compared to CDM. The bar amplitude starts to rise at $T \approx 3$ Gyr, and by the end of the simulation, the amplitude is roughly twice that of the CDM run. The $\sigma_\rmm=1 \cmg$ run shows similar evolution, except the final bar amplitude is even higher. The $\sigma_\rmm=10\cmg$ run shows drastically accelerated bar growth, reaching a peak amplitude of $\approx 0.5$ within $4$ Gyr. However, after this point, the bar amplitude decreases over time until the simulation is terminated. 

\begin{figure*}
    \centering
    \includegraphics[width=\linewidth]{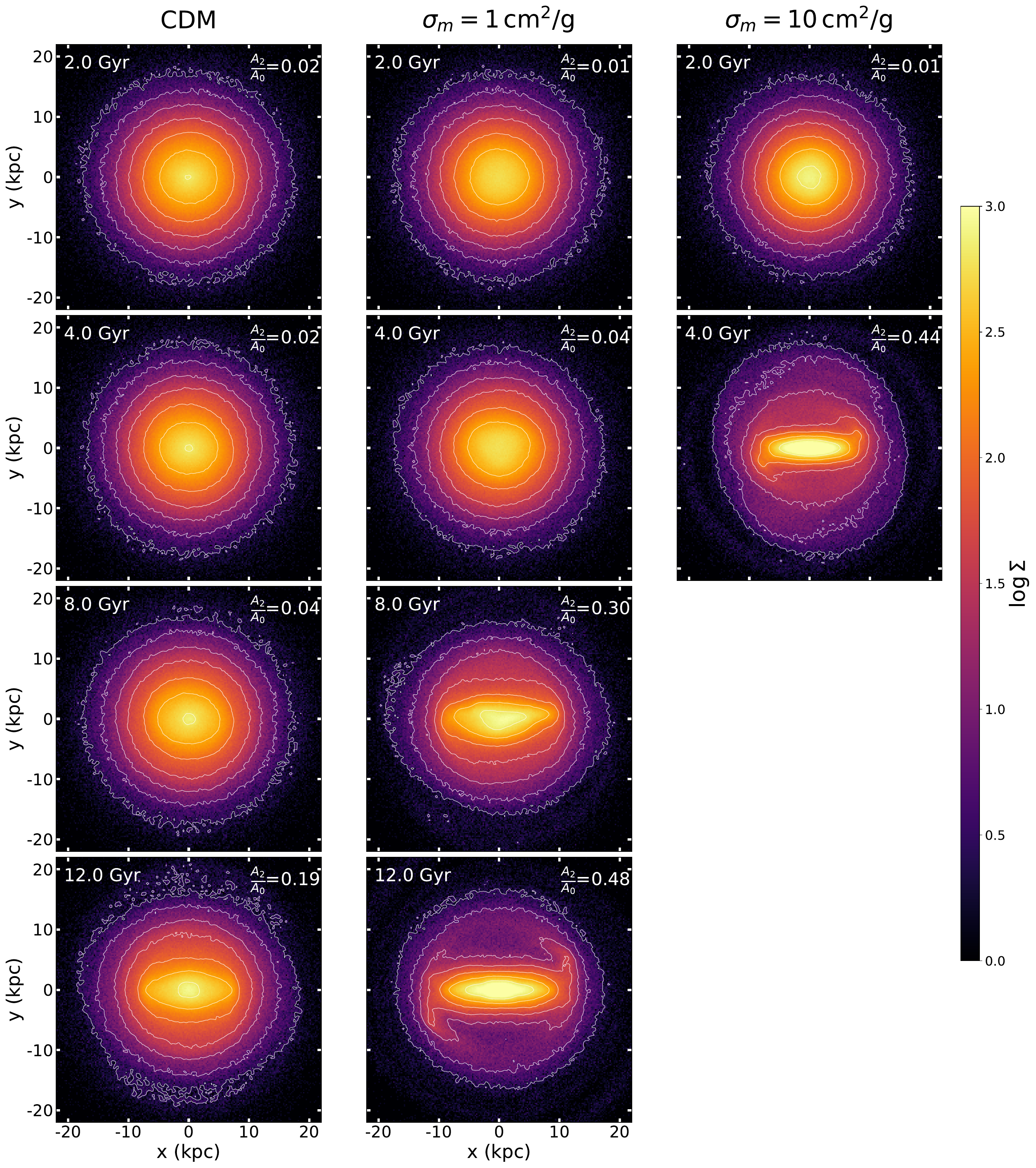}
    \caption{Projected stellar surface density in the x-y plane for the {\tt Fiducial} suite. The columns correspond to the CDM (left), $\sigma_\rmm=1\cmg$ (middle), and $\sigma_\rmm=10 \cmg$ (right) runs, with different rows corresponding to different snapshot times, as indicate in the top-left corner of each panel. The bar amplitude is also indicated, in the top-right corner. Note that the CDM run forms a moderate bar only at late times, whereas the SIDM runs form a significantly stronger bar at earlier times. The $\sigma_\rmm=10\cmg$ run is terminated at $T\approx 4.6$ Gyr once it reaches core collapse, and therefore the bottom two rows of the right-most column are empty.}
    \label{fig:snapshots}
\end{figure*}

In order to qualitatively visualize the differences between the runs, in Figure \ref{fig:snapshots} we plot the projected stellar surface density in the $x$-$y$ plane for various simulation snapshots, with the galaxy rotated such that the bar lies along the $x$-axis. From left to right, the different columns show the results for
$\sigma_\rmm=0$ (CDM), $1\cmg$, and $10 \cmg$, as indicated.
Different rows correspond to different simulation times, as indicated in the top-left corner of each panel. Since the $\sigma_\rmm=10\cmg$ run was terminated at $T\approx 4.6$ Gyr once it reached core collapse, the corresponding two bottom rows are empty. In the CDM run, a discernible bar structure is visible is not visible for $T < 8$ Gyr, and is only clearly visible in the last $12$ Gyr snapshot. On the other hand for $\sigma_\rmm=1 \cmg$, a strong bar is qualitatively visible at 8 Gyr, and by 12 Gyr it is very prominent. The $\sigma_\rmm=10 \cmg$ shows a strong, clearly identifiable bar already at 4 Gyr. 

\subsection{Trend across models}

\begin{figure*}
    \centering
    \includegraphics[width=\textwidth]{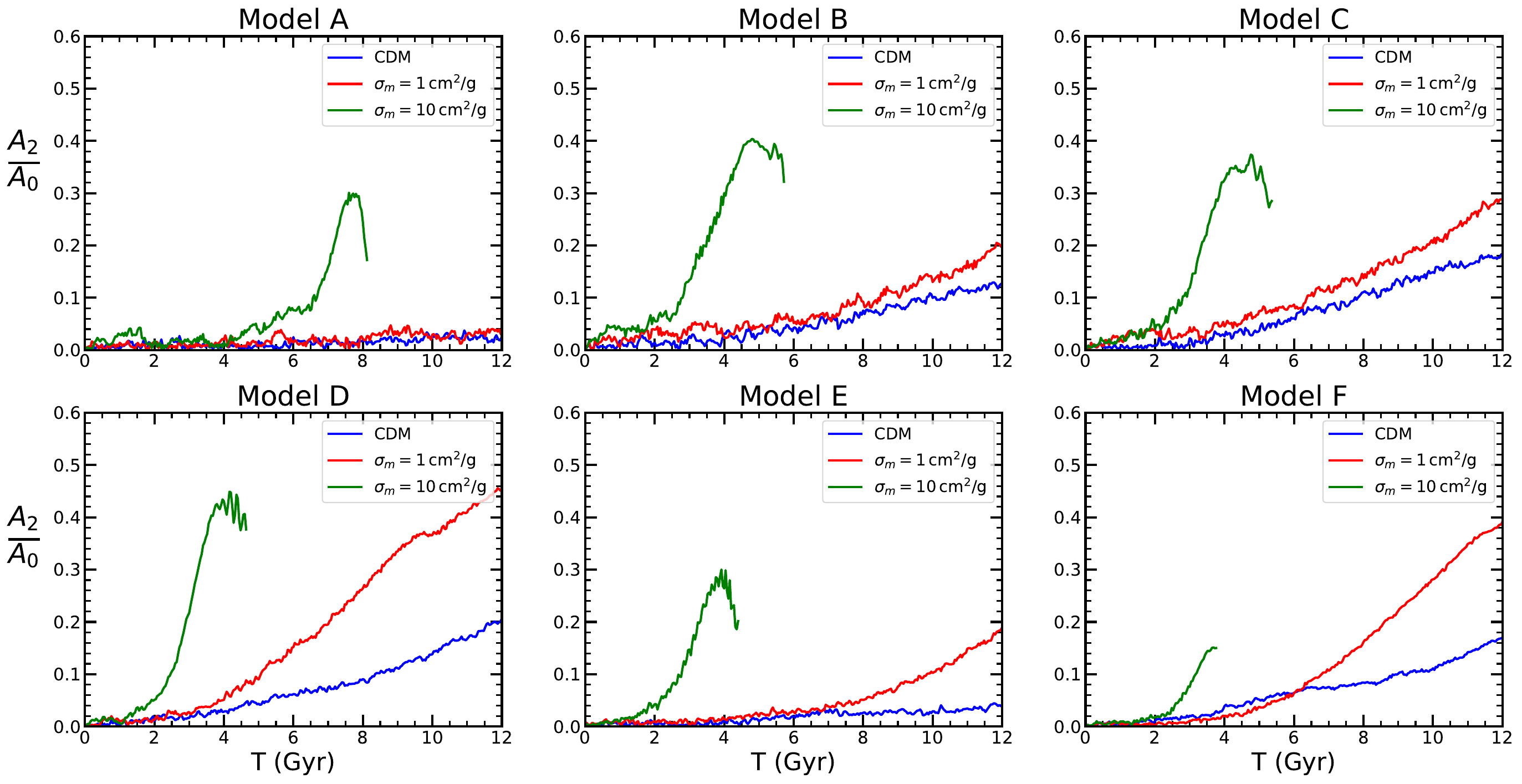}
    \caption{Bar amplitude evolution in the various models. Within each panel, the CDM, $\sigma_\rmm=1\cmg$, and $\sigma_\rmm=10\cmg$ runs are plotted. In all cases, the bar amplitude rises faster and reacher a higher value in the SIDM runs compared to CDM. Noteworthy cases are Models A and E, which are stable against bar formation in CDM, but do form bars once the cross section is dialed up.}
    \label{fig:param_space}
\end{figure*}

Figure \ref{fig:param_space} shows the bar amplitude over time for Models A-F, all of which have an initial NFW halo profile ($\gamma=1$). These models have been selected by varying $M_{\rm stellar}$ and $Q_{\rm min}$ such that bar formation occurs on a $\sim$ few Gyr timescale in CDM, which is the regime where the differences between CDM and SIDM manifest the strongest (see the discussion in Section~\ref{sec:resonances}).

Similar to the {\tt Fiducial} run, in all models the bar forms faster and reaches a higher amplitude in SIDM compared to CDM. A cross section of $\sigma_\rmm=1 \cmg$ is sufficient to noticeably accelerate bar growth in all models. When the cross section is increased to $\sigma_\rmm=10 \cmg$, the bar growth is drastically accelerated, reaching a peak amplitude within $\sim 4$ Gyr, after which it decreases in some runs. Noticeably in {\tt Model A}, which consists of a dark matter dominated disk, the CDM and $\sigma_\rmm=1\cmg$ runs do not form an appreciable bar throughout the simulation. However, when the cross section is increased to $\sigma_\rmm=10\cmg$, a strong bar forms in $\sim 7$ Gyr. Similarly in {\tt Model E}, which has a kinematically hot disk with $Q_{\rm min}=3.0$, there is no bar formation in the CDM run, a weak bar by the end of the $\sigma_\rmm=1 \cmg$ run, and a strong bar in the $\sigma_\rmm=10 \cmg$ run.

\begin{figure*}
    \centering
    \includegraphics[width=\linewidth]{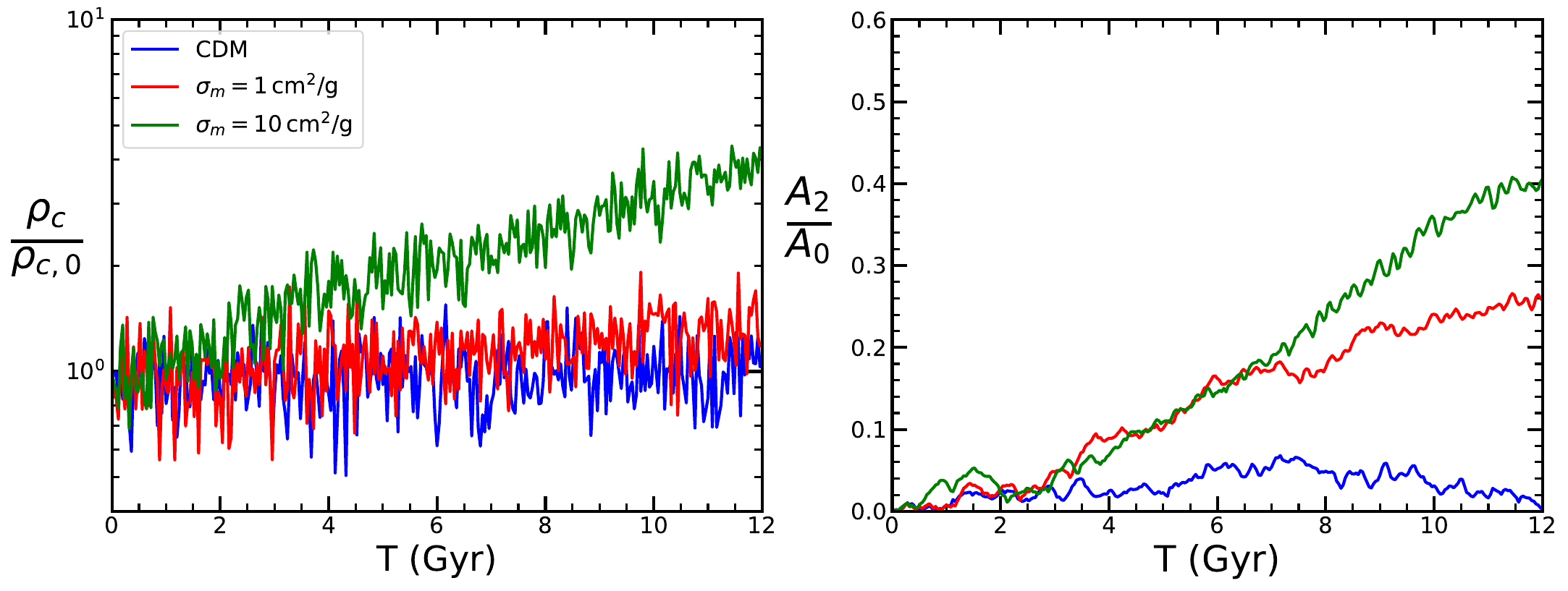}
    \caption{Normalized central dark matter density evolution (left) and bar amplitude {\it vs} time for {\tt Core} suite. As in the {\tt Fiducial} suite, the central density remains constant throughout in the CDM runs. In the SIDM runs, since the halo begins with an isothermal core, the heat conduction is always outward, resulting in a central dark matter density that continuously increases over time. Despite this, the bar amplitude trend is the same as that seen in Figures~\ref{fig:fiducial} and \ref{fig:param_space}: stronger bars are formed with increasing cross section.}
    \label{fig:bar_amp_core}
\end{figure*}

\subsection{Verification tests}
\label{ssec:verify}

From the above subsections, it is clear that bars are able to form faster and stronger in SIDM halos compared to CDM halos, when the halo starts out as an NFW profile. One might assume that this simply reflects the decreased central DM density during the core formation stage in SIDM, along the lines of previous claims that bars form more readily when the central DM density is low \citep[e.g.][]{Athanassoula2003, Reddish2022}. However, we perform two tests to show show that the accelerated bar growth in SIDM is {\it not} simply due to a reduction in DM density. The first is based on the suite of {\tt Core} simulations, in which the halo is initialized with a central constant-density core rather than a NFW-like cusp. The second is based on the suite of {\tt Freeze} simulations, in which we turn off the self-interactions after they have transformed the initial cusp into a core.

\begin{figure*}
    \centering
    \includegraphics[width=\linewidth]{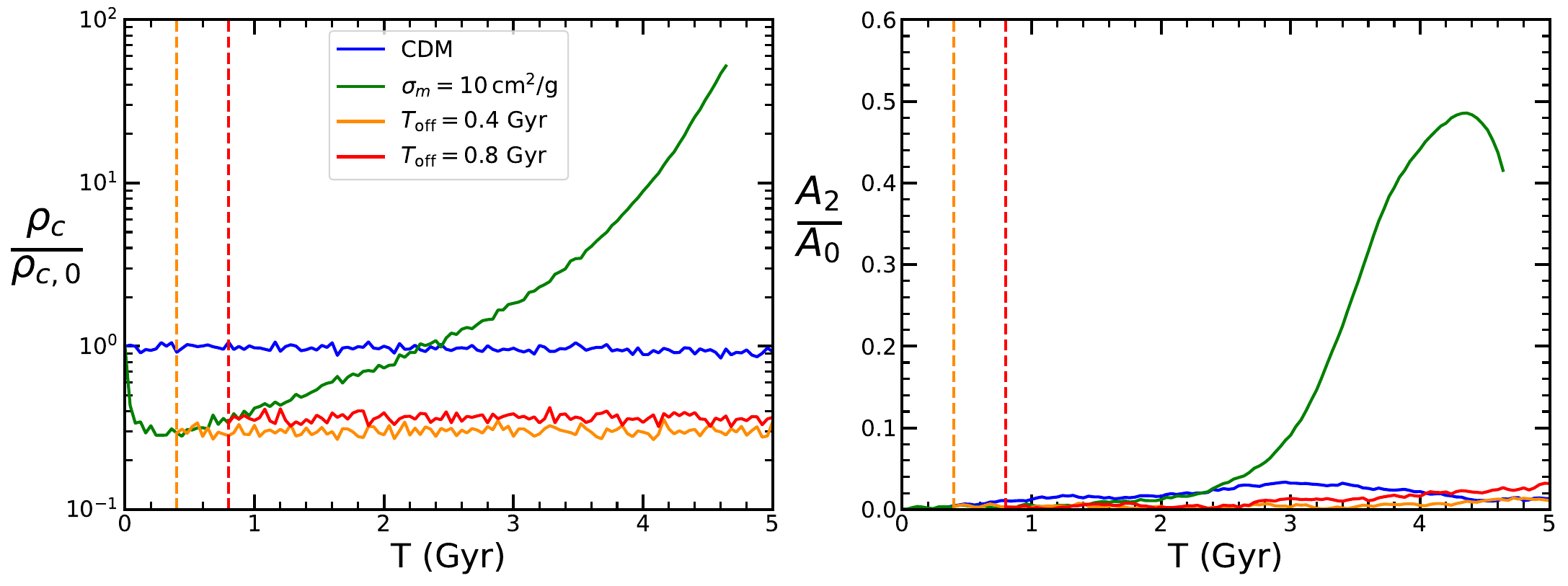}
    \caption{Normalized central dark matter density evolution (left) and bar amplitude {\it vs} time for the {\tt Fiducial} CDM (blue) and $\sigma_\rmm=10\cmg$ (green) runs, and the {\tt Freeze} $T_{\rm off}=0.4$ Gyr (orange) and $T_{\rm off}=0.8$ Gyr (red) runs. The {\tt Freeze} simulations are run by freezing the $\sigma_\rmm=10\cmg$ simulation and turning off the dark matter self-interactions at time $T_{\rm off}$, indicated by the dashed vertical lines. After this, the simulation is continued with CDM physics. Once the self-interactions are turned off, the halo is effectively frozen in its cored state, and the central density remains constant over time. Despite this, the bar amplitude remains well below the parent $\sigma_\rmm=10\cmg$ run, showing that the dark matter core formation is not the main driver of accelerated bar growth in SIDM. }
    \label{fig:bar_amp_freeze}
\end{figure*}

\subsubsection{Bar formation in cored halos}

In the suite of {\tt Core} simulations, the initial dark matter halo has an isothermal constant density core. Since the velocity dispersion of the core is constant and larger than that in the outskirts, the conductive heat transfer is always directed outward. Therefore, unlike in halos with an initial NFW profile (where the velocity dispersion profile dips towards the center), the DM density only increases over time, i.e. it never goes through a core formation stage. The left panel of Figure~\ref{fig:bar_amp_core} shows the evolution of the central DM density in the suite of {\tt Core} simulations for CDM, $\sigma_\rmm=1\cmg$, and $\sigma_\rmm=10\cmg$. In the CDM run the central density remains constant, while in the SIDM runs the density increases over time. Note that the density evolution is significantly more modest than in the {\tt Fiducial} simulation which starts out with an NFW cusp. This is because of the lower central density in the {\tt Core} halo, which results in a lower scattering rate and therefore slower heat conduction. 

The right panel of Figure~\ref{fig:bar_amp_core} shows the time evolution of the bar amplitude in the {\tt Core} simulations. We emphasize that we are not aiming to directly compare the bar properties between the {\tt Core} and other simulations, but rather to assess the impact of SIDM within the {\tt Core} suite. Similarly to Figure~\ref{fig:param_space}, the bar amplitude increases faster and/or reaches a higher peak in SIDM compared to CDM, and this effect is more pronounced for $\sigma_\rmm=10 \cmg$ compared to $1 \cmg$. Since the central DM density never decreases over time in these runs, the accelerated bar growth in SIDM clearly cannot be attributed to core formation.

\subsubsection{Turning off the self-interactions}

In the suite of {\tt Freeze} simulations, we start from the initial conditions of the {\tt Fiducial} simulation with $\sigma_\rmm=10\cmg$, but turn off the self-interactions at some time $T_{\rm off}$.
We run two different simulations with $T_{\rm off}=0.4$ Gyr and 0.8 Gyr, respectively, at which times the halo is well in its cored stage, but has not yet reached core collapse. The orange and red curves in the left panel of Figure~\ref{fig:bar_amp_freeze} show the evolution of the normalized central density for both simulations, with the dashed vertical lines indicating $T_{\rm off}$. For comparison, we also show the results for the {\tt Fiducial} SIDM simulation with the same cross section and for the {\tt Fiducial} CDM simulation. It is clear that once the self-interactions are turned off, the central density remains constant throughout the rest of the simulation, i.e. the halo is effectively frozen in its cored state.  

The right panel of Figure~\ref{fig:bar_amp_freeze} shows the amplitude of the bar $A_2/A_0$ in the same four runs. As we saw earlier, the $\sigma_\rmm=10 \cmg$ run forms a strong bar within $4$ Gyr, whereas the CDM run does not form an appreciable bar over this period. The bar amplitudes of the two {\tt Freeze} runs, which fork off of the $\sigma_\rmm=10 \cmg$ curve at 0.4 Gyr and 0.8 Gyr respectively, remain close to zero throughout, similar to that in the CDM run. Despite the presence of a core, the galaxy does not form a bar in the absence of dark matter self-interactions. Therefore, it is clear that the core formation mechanism of SIDM is not the main driver of the enhanced or accelerated bar growth.

\section{Orbital dynamics governing bar-halo interaction}
\label{sec:resonances}

We now analyze the orbital dynamics driving the bar formation and evolution, and compare the dynamics in CDM vs SIDM to understand the cause for accelerated bar growth in SIDM. 

\subsection{Brief theoretical overview}

We begin with a brief overview of the theoretical foundations for bar-halo interactions, which have been established by \citet{LBK} and \citet{TW84}, and further developed by several authors \citep[e.g.][]{Weinberg2004, Chiba2022, Chiba2023}. The key aspect is that the rotating bar mode induces a perturbation in the halo, which back-reacts on the bar. This results in a torque acting on the bar by the halo, resulting in a net angular momentum transfer from the bar to the halo. Using linear perturbation theory, \citet{LBK} and \citet{TW84} showed that the rate of angular momentum transfer from the spherical halo to the bar is governed by the LBK torque:
\begin{equation}
\label{eq:LBK}
    \frac{\rmd L_z}{\rmd t} = 8 \pi^4 \sum_{\mathbf{\ell}} l_\phi \int \rmd \mathbf{J} \left| \hat{\Phi}_\ell(\mathbf{J}) \right|^2 \mathbf{\ell} \cdot \frac{\partial f}{\partial \mathbf{J}} \delta(\mathbf{\ell} \cdot \mathbf{\Omega} - l_\phi \Omega_\rmb)\,.
\end{equation}
Here, $\mathbf{J}=(J_r,J_z,J_\phi)$ are the three actions, $\mathbf{\Omega} = (\Omega_r,\Omega_z,\Omega_\phi)$ are the corresponding frequencies, $\mathbf{\ell}=(l_r,l_z,l_\phi)$ is a vector of integers, $\Omega_\rmb$ is the pattern speed of the bar, $\left| \hat{\Phi}_l(\mathbf{J}) \right|$ are the Fourier modes of the bar potential, $\delta(x)$ is the Dirac delta function, and $f=f(J_r,J_z,J_\phi)$ is the unperturbed DF of the halo. 

It is clear from equation~(\ref{eq:LBK}) that angular momentum is only emitted or absorbed at resonances, and at each resonance, the gradient in the DF determines the direction of transfer. In other words, the net angular momentum flux is determined by the imbalance between the number of particles that gain vs. lose angular momentum across the resonance. This is a characteristic of secular transport in gravitational systems, which governs a wide variety of processes, including spiral arm formation \citep{Sellwood2022, Petersen2024, Hamilton2024}, dynamical friction \citep{Weinberg1985, Banik2021, Dattathri2025b}, and the evolution of internal gravitational modes \citep{Kalnajs1977, Heggie2020, Dattathri2025a}. In the context of bar-halo interaction, since $\partial f/\partial \mathbf{J}<0$ for most halo DFs, the net torque acting on the bar is negative, and therefore angular momentum is transferred from the bar to the halo\footnote{As pointed out by \citet{Weinberg1985}, angular momentum can also be internally redistributed within the galaxy due to bar-disk resonances. The bar emits angular momentum at the inner Lindblad resonance, which is absorbed by the outer disk at the corotation and outer Lindblad resonances \citep{Athanassoula2003}. However, the halo absorbs more angular momentum since is is significantly more massive than the disk, making it the primary focus of this work.}. 

\subsection{Effect of collisions}
\label{ssec:theory}

Equation~(\ref{eq:LBK}) is derived from the collisionless Boltzmann equation, and therefore is only applicable in CDM. As shown in Appendix~\ref{app:lbk}, in the presence of collisions, as in the case of SIDM, the torque responsible for the angular momentum transfer is given by
\begin{equation}
\begin{split}
\label{eq:SIDM_LBK}
\frac{\rmd L_z}{\rmd t} = 8 \pi^3 \sum_{\mathbf{\ell}} l_\phi \int \rmd \mathbf{J} \,\left| \hat{\Phi}_\ell(\mathbf{J}) \right|^2 \, \mathbf{\ell} \cdot \frac{\partial f}{\partial \mathbf{J}} \\
\times \frac{\nu(\mathbf{J})}{(\mathbf{\ell} \cdot \mathbf{\Omega} - l_\phi \Omega_\rmb)^2 +\nu(\mathbf{J})^2}   \,,
\end{split}
\end{equation}

where $\nu(\mathbf{J})$ is the orbit-averaged collision timescale of a particle with action $\mathbf{J}$. 

Equation~(\ref{eq:SIDM_LBK}) indicates that collisions {\it broaden} the resonances, replacing the Dirac delta function of Equation~(\ref{eq:LBK}) with a Lorentzian of half-width $\nu(\mathbf{J})$\footnote{The Dirac delta function in Equation~(\ref{eq:LBK}) is a direct mathematical consequence of taking the time-asymptotic limit (see Appendix~\ref{app:lbk}), i.e. the bar evolution timescale is assumed to be infinitely longer than the orbital timescale. If this limit is relaxed, resonance broadening is possible within CDM as well, as shown by \citet{Weinberg2004, Banik2021}. In such a case, collisions due to SIDM will {\it further} broaden the resonances. In this paper, ``resonance broadening" refers to the broadening in SIDM compared to CDM, regardless of whether the CDM resonance is a strict delta function or not.}. Therefore, a larger volume of phase-space is able to couple to the bar and contribute to the torque. Physically, this represents the fact that collisions can scatter DM particles, which otherwise would not interact with the bar, into the vicinity of a resonance such that they can do so. With increasing SIDM cross section, the phase-space response is further broadened, giving rise to a larger torque. Note that in the limit $\nu(\mathbf{J}) \rightarrow 0$ (i.e. the collisionless limit), the Lorentzian asymptotes to a delta function:
\begin{equation}
    \lim_{\nu(\mathbf{J}) \to 0}\frac{\nu(\mathbf{J})}{(\mathbf{\ell} \cdot \mathbf{\Omega} - l_\phi \Omega_\rmb)^2 +\nu(\mathbf{J})^2} = \pi \delta(\mathbf{\Omega} - l_\phi \Omega_\rmb)  \,,
\end{equation}
and the standard LBK torque formula is recovered.

To demonstrate the resonance broadening and its effect on bar-halo torque, we consider the $T=12$ Gyr snapshot of the {\tt Fiducial} CDM simulation, and calculate the torque from the corotation resonance as predicted by Equations~(\ref{eq:LBK}) and (\ref{eq:SIDM_LBK}). The full details of the calculation are given in Appendix~\ref{ssec:torque_calculation}. We find that the corotation torque values, in units of $M_\odot \, {\rm kpc^2 \, Gyr^{-2}}$, are: $-5.67 \times 10^{11}$, $-7.73 \times 10^{11}$, and $-1.53 \times 10^{12}$ for CDM, $\sigma_\rmm=1 \cmg$, and $\sigma_\rmm=10\cmg$ respectively. Therefore, for a fixed bar amplitude, increasing the cross section from $0$ to $10 \cmg$ results in a factor of $\sim 2.7$ increase in the corotation torque. 
    
Although the LBK torque provides powerful insight, it is based on linear perturbation theory and therefore does not take into account non-linear effects such as orbit trapping. Hence, the LBK torque and its SIDM equivalent are only applicable in the ``fast regime'', where the pattern speed of the bar is changing fast enough such that orbit trapping of dark matter by the bar is negligible. In Appendix~\ref{ssec:pendulum} we use the pendulum approximation to show that all bars in our simulations are indeed in the fast regime. 

\subsection{Angular momentum transfer between the disk and halo}

\begin{figure}
    \centering
    \includegraphics[width=\linewidth]{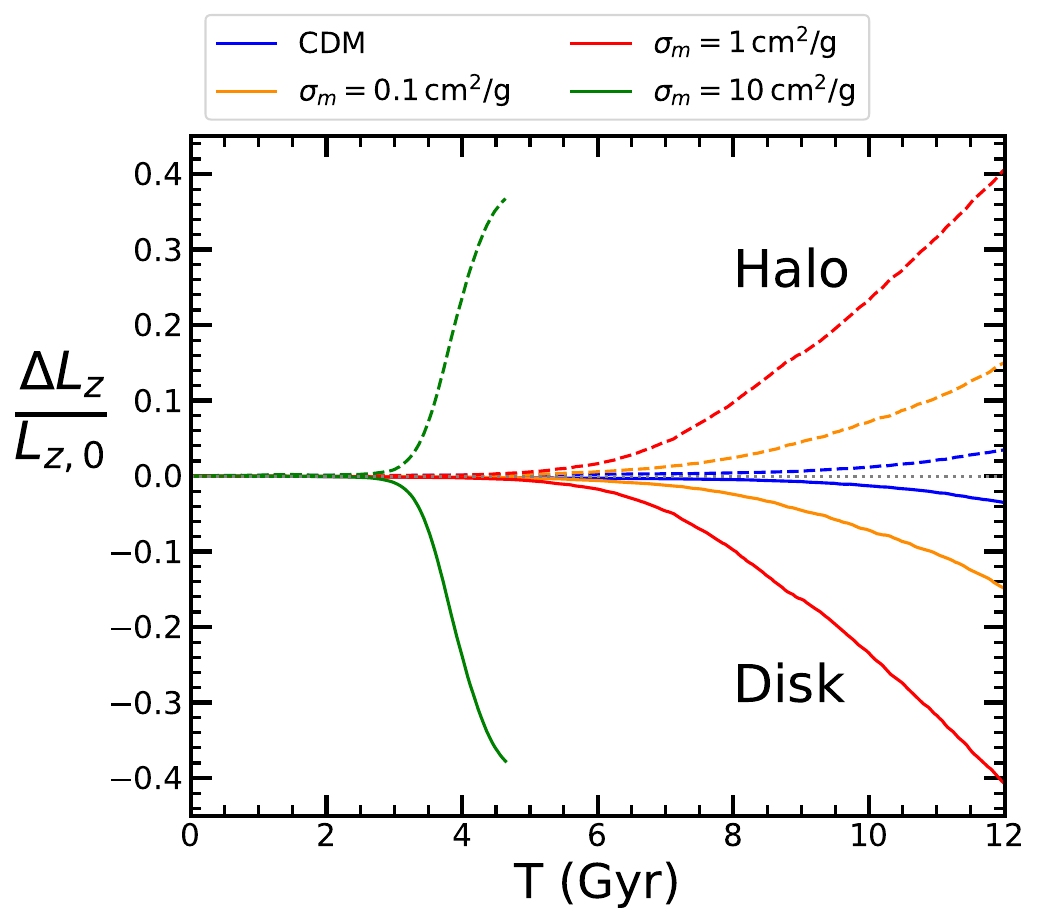}
    \caption{Change in the total z-component angular momentum $\Delta L_z$ of the disk (solid lines) and halo (dashed lines) in the {\tt Fiducial} suite, normalized by the disk's initial value $L_{z,0}$. In all runs, the disk transfers angular momentum to the halo, and this transfer is more efficient in SIDM, and with increasing cross section, than in CDM.}
    \label{fig:disk_halo_lz}
\end{figure}

First, we compare the overall angular momentum evolution of the disk and halo in the various runs. In Figure \ref{fig:disk_halo_lz}, we plot the change in the $z$-component of the angular momentum ($\Delta L_z=L_z(t)-L_{z,0}$) normalized by the disk's initial angular momentum ($L_{z,0}$) for the {\tt Fiducial} suite of simulations. In every case, over time the disk loses angular momentum while the halo gains the same amount (down to numerical accuracy). This reflects the disk shedding its angular momentum, driving bar growth. 

It is clear from Figure \ref{fig:disk_halo_lz} that the angular momentum transfer from the disk to the halo is significantly more efficient in SIDM compared to CDM, and increasingly more efficient for higher cross section. Even with a cross section as low as $\sigma_\rmm=0.1\cmg$, by the end of the simulation, the disk has lost over four times the angular momentum as in the CDM case.

\subsection{Phase-space resonance structure}
\label{ssec:phase_space}
\begin{figure*}
    \centering
    \includegraphics[width=0.495\linewidth, trim= 50 0 37 0]{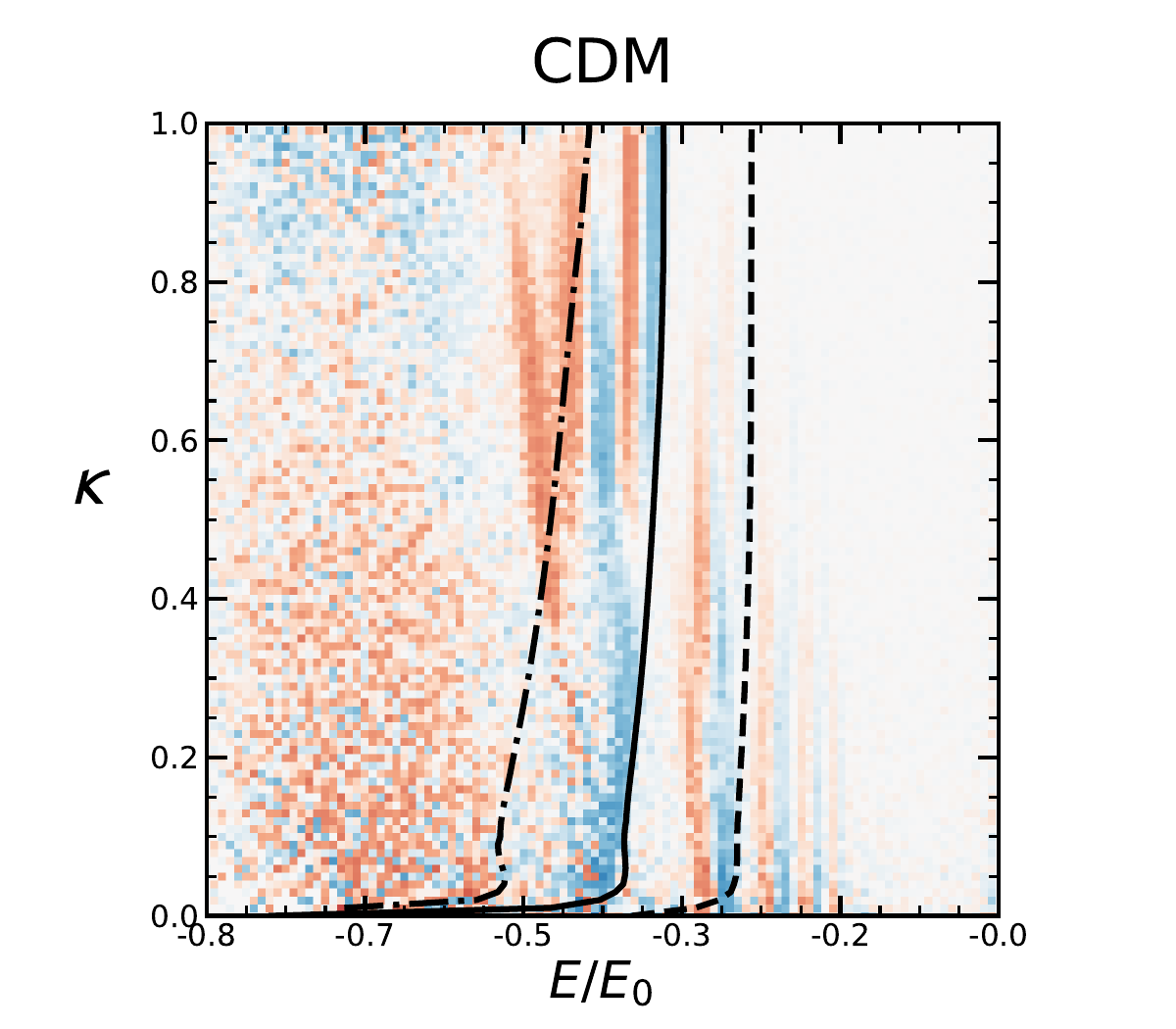}
    \includegraphics[width=0.495\linewidth, trim= 37 0 50 0]{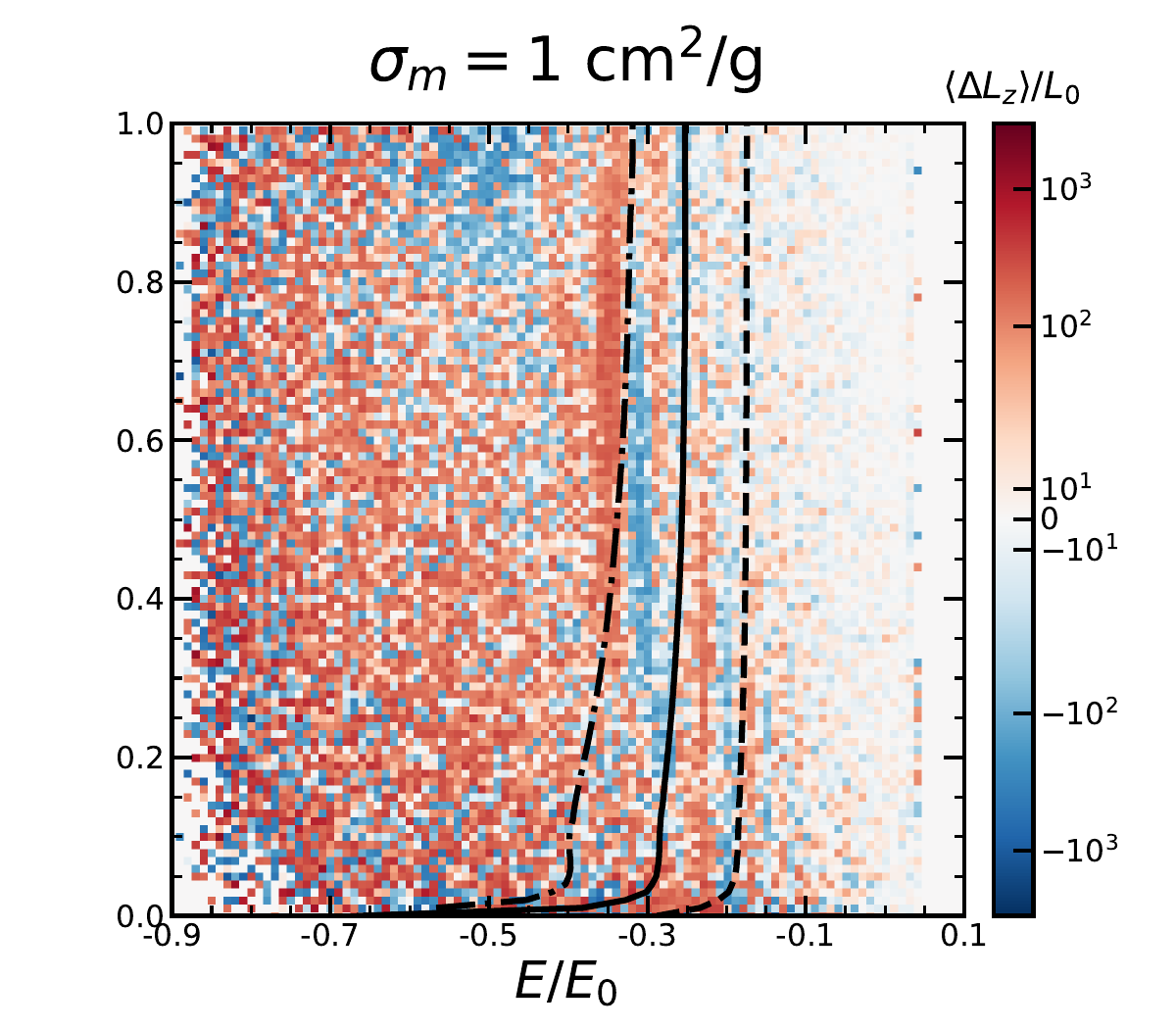}
    
    \caption{Change in the halo's $z$-component of angular momentum, $\langle \Delta L_z \rangle$, between $T=10$ Gyr and 12 Gyr, binned in the $E$ and $\kappa \equiv L/L_{\rm c}(E)$ plane. Both $E$ and $\langle \Delta L_z \rangle$ are in units of $E_0=1 \, {\rm kpc^2 \, Gyr^{-2}}$ and $L_0=1 \, {\rm kpc \, km \ s^{-1}}$. The left and right panels correspond to the CDM run and to the SIDM run with $\sigma_\rmm=1 \cmg$, respectively. The black lines represent resonance loci, corresponding to the ultraharmonic resonance (dotted-dashed), the corotation resonance (solid), and the outer Lindblad resonance (dashed). In the CDM run, the $\Delta L_z$ features closely surround the resonances, whereas in the $\sigma_\rmm=1\cmg$ run they are stronger and less clearly structured. See text for discussion.}
    \label{fig:phase_space}
\end{figure*}

We now map the angular momentum changes seen above onto the phase-space structure of the system. As seen in equation~(\ref{eq:LBK}), secular theory predicts that the action (or angular momentum) flux is transferred only at resonances, at least in collisionless systems. In an axisymmetric disk+halo system, resonances are defined by the commensurability condition
\begin{equation}
    l_r \Omega_\rmr + l_\phi \Omega_\phi - l_\rmb \Omega_\rmb=0 \ , 
\end{equation}
where $\Omega_r$ and $\Omega_\phi$ are the radial and azimuthal frequencies, and as before, $l_r,l_\phi,l_\rmb$ are integers. We first construct an axisymmetric \texttt{Multipole} potential from the N-body snapshot,
and then compute the frequencies using the St{\"a}ckel approximation \citep{Binney2012}, as implemented in \texttt{Agama}. The frequencies are tabulated in the $E-\kappa$ plane, where $E$ is energy and $\kappa=L/L_{\rm c}(E)$ is the angular momentum normalized by the corresponding value of a circular orbit with the same energy. These provide a convenient rectangular coordinate system to visualize the phase-space structure of the system. The pattern speed of the bar $(\Omega_\rmb)$ is estimated by measuring the rotation rate of the disk's moment of inertia tensor. 

In the left panel of Figure~\ref{fig:phase_space} we plot the change in the $z$-component of the angular momentum $\Delta L_z$ of the halo particles in the CDM run, between $T=$ 10 Gyr and 12 Gyr\footnote{The bar pattern speed is continuously decreasing due to dynamical friction against the halo. However, over this time period, it is approximated to be constant.}, binned in the $E-\kappa$ plane. Note how the $\Delta L_z$ values are not randomly distributed, but rather are highly structured. The positive and negative $\Delta L_z$ regions occur around the resonance loci, three of which are overlaid on the plot: the co-rotation resonance $(0,1,1)$, the outer Lindblad resonance $(1,2,2)$, and the ultra-harmonic resonance $(-1,4,4)$. These features represent the halo particles that have gained or lost angular momentum at that resonance. This is evident from equation~(\ref{eq:LBK}): the orbit response on either side of a resonance is equal but opposite in sign.  

In the right panel of Figure~\ref{fig:phase_space} we show the same, but for the $\sigma_\rmm=1\cmg$ run. Compared to the CDM case, the positive and negative $\Delta L_z$ features are stronger and less clearly structured. This is because collisionality due to self-interactions aids in the angular momentum transfer by scattering particles close enough to a resonance such that they can sweep across it (i.e. the resonances are effectively broadened). Collisions occur stochastically in phase-space instead of along resonances, and therefore work towards erasing the structured $\Delta L_z$ features seen in the case of CDM. In the $\sigma_\rmm=1\cmg$ case, the three main resonances are still qualitatively visible, but the higher order resonances are no longer visible. 

\subsection{Bar destruction at late-stage core collapse}

\begin{figure}
    \centering
    \includegraphics[width=\linewidth]{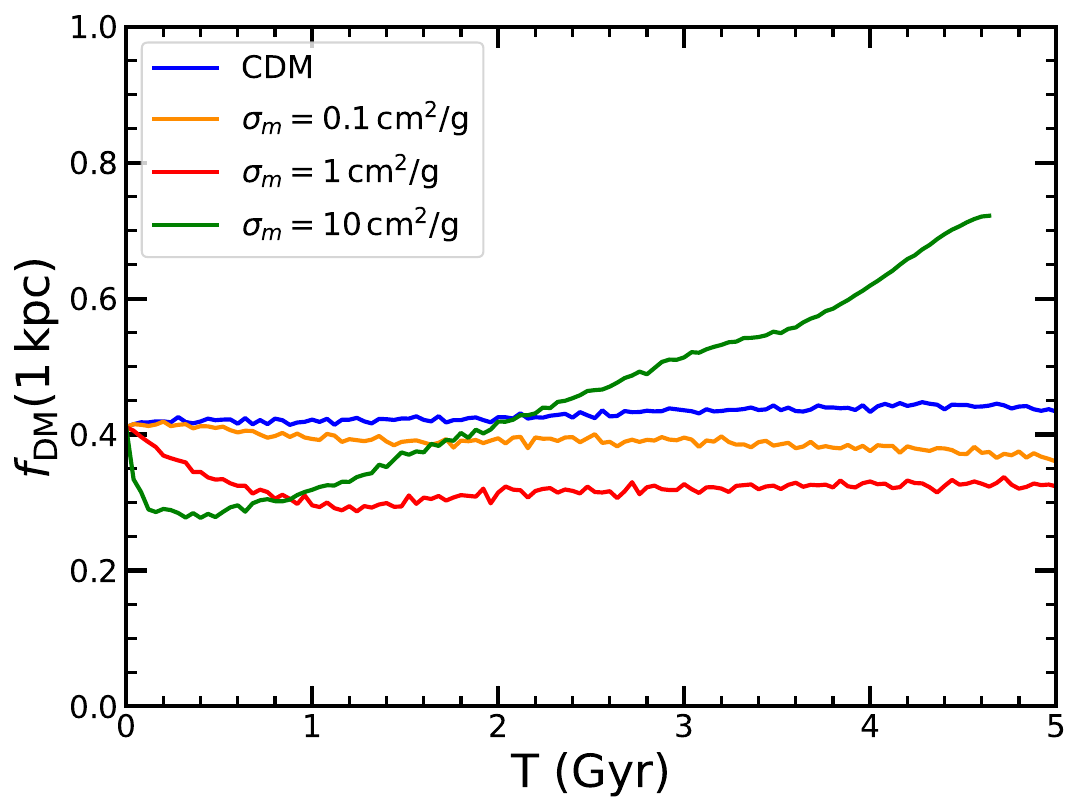}
    \caption{Enclosed dark matter fraction within 1 kpc over time in the {\tt Fiducial} suite. In the CDM run, $f_{\rm DM}$(1 kpc) remains constant over time, whereas in the SIDM $\sigma_\rmm=0.1 \cmg$ and $\sigma_\rmm=1\cmg$ runs it decreases out to 5 Gyr due to dark matter core formation. However in the $\sigma_\rmm=10 \cmg$ run, the halo quickly enters core collapse, and the value of $f_{\rm DM}$(1 kpc) increases over time after the first $\sim 0.5$ Gyr. By the end of the simulation, the galaxy is highly dark matter dominated, resulting in bar weakening and reduction in its amplitude (see Figure~\ref{fig:fiducial}).}
    \label{fig:fdm}
\end{figure}

While the $\sigma_\rmm=10 \cmg$ runs in all models show significantly faster bar growth, towards the end of the simulations the bar amplitudes start to decline. To understand the cause for this, in Figure \ref{fig:fdm} we plot the enclosed dark matter fraction within 1 kpc (i.e. $f_{\rm DM} (1 \, {\rm kpc}) = M_{\rm DM,enc}(1 \, {\rm kpc})/M_{\rm tot,enc}(1 \, {\rm kpc})$) over time for the {\tt Fiducial} suite. As expected, in CDM, $f_{\rm DM}$ remains constant over time since the DM halo does not significantly evolve (see Figure~\ref{fig:fiducial}). In the SIDM runs, the value of $f_{\rm DM}$ first drops, due to the dark matter core formation, and then slowly increases over time as the halo heads toward core collapse. In the $\sigma_\rmm=0.1 \cmg$ and $1 \cmg$ runs, the core collapse takes much longer than a Hubble time, so the value of $f_{\rm DM}$ always remains $\leq 0.4$, i.e. the galaxy is always baryon dominated. 

However, the $\sigma_\rmm=10 \cmg$ run rapidly approaches core collapse, which is reflected in the evolution of $f_{\rm DM}$. At the late stages of core collapse ($T \geq 4$ Gyr), the central region of the galaxy becomes dark matter dominated with a value $f_{\rm DM} \geq 0.6$. Such dark matter dominated disks are stable against bar instability \citep{Debattista2000, Shen2004}. In addition, the high central DM density weakens the bar by destroying the bar-supporting ($x_1$) orbits, akin to bar destruction by central mass concentrations \citep{Norman1996, Hozumi1998, Shen2004}. 

Note that in our simulation, we are able to model the bar amplitude reduction only for $\sim 0.5$ Gyr before the simulation is terminated. Our code cannot simulate late-stage core collapse, as the simulation timesteps become increasingly small as the central density increases. We expect that if the run were continued, the bar amplitude would drop further until complete dissolution. However, verifying this is beyond the scope of this paper. 

\section{Discussion and conclusions}
\label{sec:conclusion}

Over half a century of in-depth studies have revealed that bar formation and evolution is a complex dynamical process depending on several factors, including but not limited to disk self-gravity, kinematic temperature, central mass concentrations, and the live response of the dark matter halo. Particularly relevant for this paper is the dynamic coupling between bars and their host halos: the rotating non-axisymmetric perturbation associated with the bar torques the halo, while the halo response back-reacts on the bar and controls its growth, slowdown, and long-term survival. Analytic calculations (based on linear perturbation theory and/or non-linear approximations), idealized simulations, and full cosmological simulations have all been instrumental in advancing our understanding of bar dynamics (see references in Section~\ref{sec:intro}). However, the complicated interplay between all these various factors limits our predictive power for bar strength and growth rate across all environments. This is especially true once the dark matter is allowed to depart from its collisionless CDM limit.

In this paper, we have studied the influence of dark matter self-interactions on bar formation and evolutions. Our main findings are as follows:
\begin{enumerate}
    \item In all the models explored here, disk galaxies embedded within SIDM halos form bars faster and with greater amplitude than their CDM counterparts. This is highly sensitive to the SIDM cross section, with values as small as $\sigma_\rmm=0.1 \cmg$ yielding significant differences in the final bar amplitude for Milky-Way analogues (Figure~\ref{fig:fiducial}). For $\sigma_\rmm=10 \cmg$, bar formation occurs very quickly, often within $\sim 3$ Gyr. More strikingly, SIDM can also shift the effective threshold for bar instability. Kinematically hot or dark matter dominated disks, which are stable against bar formation within a Hubble time in CDM, can go bar unstable within a few Gyr in SIDM if the cross section is sufficiently high (Figure~\ref{fig:param_space}). 
    
    \item The reason for this accelerated bar growth is not simply due to dark matter core formation, but rather because SIDM allows for a greater rate of angular momentum transfer between the disk and the halo (Figure~\ref{fig:disk_halo_lz}). In CDM, the angular momentum flux transfer occurs strictly at resonances, such that only the particles close to the resonances are able to gain or lose angular momentum. However, self-interactions or collisions between the dark matter particles broaden the resonances, coupling a larger phase-space volume of the halo to the rotating bar (Figure~\ref{fig:phase_space}). This allows the disk to shed off its angular momentum faster, leading to faster bar growth.
    
    \item Once the halo enters core collapse, the inner galaxy becomes dark matter dominated (Figure~\ref{fig:fdm}). This destroys the bar-supporting orbits, leading to a decline in bar amplitude at late times. 
\end{enumerate}

These findings have a number of important implications. Most directly, they suggest that the barred galaxy population statistics may provide a new and complementary route to testing the particle nature of dark matter. Based on these results, we predict that the fraction of barred galaxies across time and the distribution of their amplitudes should increase with increasing cross section (as long as core collapse has not yet occurred). Since SIDM can shift the effective bar instability threshold, barred galaxies that are observed to be dispersion or dark matter-dominated, which are thought to be stable against bar formation in CDM, would be especially valuable in this regard. Upcoming wide-field survey telescopes such as LSST, Euclid, and the Roman Space Telescope will enable these measurements in a wide redshift range.

In addition, CDM and SIDM make distinct predictions for bar longevity. In CDM, the bars that do form persist until the end of the simulation, and hence bar survival is primarily limited by whether the disk forms a bar at all. On the other hand in SIDM, early bar formation can be followed by late-time weakening or dissolution if core collapse proceeds sufficiently far. Bar persistence therefore encodes both the efficiency of disk-halo angular momentum coupling and the gravothermal age of the halo. The presence of bars (especially in dwarf galaxies) may be used to rule out cross sections in which core collapse would have already occurred. 
 
Bars are now being detected in substantial numbers at high redshift with JWST, including at epochs where disks are expected to be dynamically young, pointing toward rapid and efficient bar formation \citep{Erwin2018,Guo2023, Costantin2023,LeConte2024,Guo2025,Geron2025,LeConte2026, Ivanov2026}. It is also noteworthy that high-redshift disks are often not dynamically cold or well-settled, but rather dispersion-dominated or irregular \citep{Genzel2006,Law2009,Kassin2012,Wang2024}, which are thought to be non-conducive to bar formation. Interestingly, all the models presented here with $\sigma_\rmm=10 \cmg$ quickly go bar unstable within $\sim$ few Gyr. Hence, taking these results at face value, the JWST observations therefore favor a non-zero DM self-interaction cross section. However, alternative explanations within CDM remain viable, including baryon-dominated disks \citep{BlandHawthorn2023}, radial gas inflow \citep{BlandHawthorn2024}, and high-spin DM halos \citep{Kataria2025}. 

Finally, the connection between bars and central gas transport makes this dynamical effect potentially relevant for black hole fueling. Large-scale bars exert gravitational torques that remove angular momentum from gas and drive inflow from kiloparsec scales to the central few hundred parsecs, where subsequent processes such as nuclear spirals, secondary bars, gravitational instabilities, turbulence, or viscous torques can continue the transport toward the accretion disk. Thus, dark matter physics that changes the incidence, strength, and formation epoch of bars may also alter the timing and duty cycle of nuclear gas supply, central star formation, nuclear star cluster growth, and supermassive black hole accretion.

Several caveats should be emphasized. The simulations presented here are idealized and collisionless in the baryonic component, with no gas, star formation, satellite perturbations, or mergers. These ingredients can modify bar growth, slowdown, and survival \citep{Athanassoula2013,Zana2019, Beane2023, BlandHawthorn2024,Weinberg2025}, and they will also interact non-trivially with SIDM gravothermal evolution \citep{Elbert2018, Sameie2021, Silverman2026}. In addition, we have only explored a very limited parameter space, since all the scale lengths of the disk, bulge, and halo were kept constant, as was the halo mass. A fully self-consistent cosmological treatment with hydrodynamics is therefore needed before directly comparing with observations. Nevertheless, this study represents the first analysis of bar dynamics in an SIDM framework, providing a foundation for future investigations incorporating more realistic physics and cosmological environments.

\section*{Acknowledgements}

We thank Rimpei Chiba, Chris Hamilton, Monica Valluri, Behzad Tahmasebzadeh, Annika Peter, Kathryn Johnston, and Lucio Mayer for useful discussions. We thank the authors of the {\tt nanoflann} library \citep{blanco2014nanoflann} for making their code open source, which was used for the nearest neighbors calculation in the simulations. SD and FvdB are supported by the NSF through grant AST2407063. P.N. also acknowledges support from the John Templeton Foundation via Grant \#126613 and support from the Department of Energy via the grant DE-SC0017660. EV acknowledges support from an STFC Ernest Rutherford fellowship (ST/X004066/1).

\section*{Data Availability}

The data underlying this article will be shared on reasonable
request to the corresponding author.



\bibliographystyle{mnras}
\bibliography{references} 




\appendix

\section{Bar-halo torque with collisions}
\label{app:lbk}

The DF of the halo obeys the Boltzmann equation:
\begin{equation}
    \frac{\partial f}{\partial t} + \left\{ f,H\right\} = C\left[f\right] \ ,
\end{equation} 
where $H$ is the Hamiltonian, $C[f]$ is the collision operator, and the curly brackets represent the Poisson bracket in action-angle space:
\begin{equation}
\label{eq:col_BE}
    \left\{ f,H\right\} = \sum_i \frac{\partial f}{\partial \theta_i} \frac{\partial H}{\partial J_i} - \frac{\partial f}{\partial J_i} \frac{\partial H}{\partial \theta_i}  \ . 
\end{equation}

The DF and the Hamiltonian are perturbed to linear order:
\begin{equation}
\label{eq:pert}
    f = f_0 + f_1 \quad \quad H = H_0 + \Phi_1 \ , 
\end{equation}
where $f_0$ and $H_0$ are the unperturbed DF and Hamiltonian of the system respectively, $f_1$ is the response, and $\Phi_1$ is the potential of the bar. We assume that the unperturbed DF describes the system in collisionless equilibrium so it obeys the Jeans theorem, i.e. $f_0 =f_0(\mathbf{J})$. In reality, $f_0$ itself evolves over time due to collisions. However, our assumption remains valid for sufficiently small SIDM cross sections for which the collisional relaxation time is long compared to the dynamical time.

The perturbed DF and bar potential are expanded as a Fourier series in action-angle variables:
\begin{equation}
\label{eq:f1}
    f_1 = \sum_\mathbf{\ell} f_{1,\mathbf{\ell}}(\mathbf{J},t) e^{i \mathbf{\ell}\cdot\boldsymbol{\theta}}
\end{equation}
\begin{equation}
\label{eq:phi1}
    \Phi_1 = \sum_\mathbf{\ell} \Phi_{1,\mathbf{\ell}}(\mathbf{J},t) e^{i \mathbf{\ell}\cdot\boldsymbol{\theta}} \ . 
\end{equation}

The total torque on the halo is given by (cf., Equation~[20] in \citealt{Banik2021}):
\begin{equation}
    \frac{\rmd L_z}{\rmd t} = \int \rmd\boldsymbol{\theta} \, \rmd\mathbf{J} \, f_1 \, \frac{\partial \Phi_1}{\partial \phi} \ . 
\end{equation}
Substituting the above expressions for $f_1$ and $\Phi_1$ yields
\begin{equation}
    \frac{\rmd L_z}{\rmd t} = 8 \pi^3 \sum_\mathbf{\ell} \int \rmd \mathbf{J} \ l_\phi \ {\rm Im} \left[ \Phi^*_{1,\mathbf{\ell}}(\mathbf{J},t) \ f_{1,\mathbf{\ell}}(\mathbf{J},t) \right] \ , \label{eq:torque1}
\end{equation}
where we have used $\int \rmd \boldsymbol{\theta} e^{i(\mathbf{\ell}-\mathbf{\ell}')\boldsymbol{\theta}}=8\pi^3 \delta_{\mathbf{\ell},\mathbf{\ell'}}$ in three dimensions and $\Phi_{1,\mathbf{-\ell}}=\Phi^*_{1,\mathbf{\ell}}$.

In order to calculate $f_{1,\mathbf{\ell}}$, we substitute Equation~(\ref{eq:pert}) into Equation~(\ref{eq:col_BE}) and expand the Poisson bracket. Keeping only the linear order terms then gives
\begin{equation}
\label{eq:linear_be}
  \frac{\partial f_1}{\partial t} + \mathbf{\Omega} \cdot \frac{\partial f_1}{\partial \boldsymbol{\theta}} -\frac{\partial \Phi_1}{\partial \boldsymbol{\theta}} \frac{\partial f_0}{\partial \mathbf{J}}= C\left[f\right] \ , 
\end{equation}
where we have used $\mathbf{\Omega}=\partial H_0/\partial \mathbf{J}$.

We now model $C[f]$ using the BGK operator \citep{Bhatnagar1954}:
\begin{equation}
    C[f] = -\nu (f - f_{\rm eq}) \ , 
\end{equation}
where $\nu = \nu(\mathbf{J})$ is the collision frequency for a particle with action $\mathbf{J}$ and $f_{\rm eq}$ is the late-time equilibrium DF (not to be confused with $f_0$, the system's DF in {\it collisionless} equilibrium). Collisions will initially drive the system towards an isothermal core, which has $f \propto e^{-E/\sigma^2}$, after which the system undergoes core collapse. The timescale for core collapse is a few hundred collision times \citep{Balberg.etal.02, Koda.Shapiro.11}. On the much shorter collisional timescale, we can ignore the evolution in $f_0$ and apply the linear Krook approximation (e.g. \citealt{Stix1962}):
\begin{equation}
\label{eq:krook}
    C_{\rm lin}[f] \approx -\nu f_1 \ .
\end{equation}

Substituting equations (\ref{eq:f1}), (\ref{eq:phi1}), and (\ref{eq:krook}) into equation (\ref{eq:linear_be}) and taking the Fourier transform, we obtain
\begin{equation}
    \frac{\partial f_{1,\mathbf{\ell}}(\mathbf{J},t)}{\partial t} + i \mathbf{\ell}\cdot \mathbf{\Omega} f_{1,\mathbf{\ell}}(\mathbf{J},t) - i \mathbf{\ell}\cdot \frac{\partial f_0}{\partial \mathbf{J}} \Phi_{1,\mathbf{\ell}}(\mathbf{J},t) = - \nu f_{1,\mathbf{\ell}}(\mathbf{J},t) \,,
\end{equation}
which can be solved using the Green's function technique with initial condition $f_{1,\mathbf{\ell}}(\mathbf{J},0)=0$ to obtain:
\begin{equation}
\label{eq:f1l}
    f_{1,\mathbf{\ell}}(\mathbf{J},t) = i \mathbf{\ell} \cdot \frac{\partial f_0}{\partial \mathbf{J}} \int_0^t \Phi_{1,\mathbf{\ell}}(\mathbf{J},t') e^{(i \mathbf{\ell}\cdot \mathbf{\Omega}+\nu)t'} \rmd t' \ .
\end{equation}
The time dependence of the bar potential can be expressed as $\Phi_{1,\mathbf{\ell}}(\mathbf{J},t)=\Phi_{1,\mathbf{\ell}}(\mathbf{J})e^{-i l_\phi \Omega_\rmb t}$. Substituting this into Equation~(\ref{eq:f1l}) and performing the integration yields
\begin{equation}
    f_{1,\mathbf{\ell}}(\mathbf{J},t) = \frac{\mathbf{\ell} \cdot \frac{\partial f_0}{\partial \mathbf{J}} \Phi_{1,\mathbf{\ell}}(\mathbf{J},t)}{\Delta \Omega + i \nu} \left[ e^{\left( i\Delta \Omega + \nu \right)t}-1 \right]\,,
\end{equation}
where we have introduced $\Delta \Omega = \mathbf{\ell}\cdot \mathbf{\Omega} - l_\phi \Omega_\phi$ for short. 

Substituting this expression for $f_{1,\mathbf{\ell}}(\mathbf{J},t)$ into Equation~(\ref{eq:torque1}) gives:
\begin{equation}
\label{eq:timedep_sidm_lbk}
\begin{split}
    \frac{\rmd L_z}{ \rmd t} = 8 \pi^3 \sum_{\mathbf{\ell}} \int \rmd \mathbf{J} \ l_\phi \ \left| \Phi_{1,\mathbf{\ell}} (\mathbf{J},t) \right|^2 \ \mathbf{\ell} \cdot \frac{\partial f_0}{\partial \mathbf{J}}\\ \times \frac{\left[ \nu - e^{-\nu t} \left( \nu \cos(\Delta \Omega t)  - \Delta \Omega \sin(\Delta \Omega t) \right) \right]}{ \Delta \Omega^2+ \nu^2} \ .
\end{split}
\end{equation}
The second term within the square brackets represents the transient response which decays exponentially over time due to collisions. Therefore, in the time-asymptotic limit ($t \rightarrow \infty$), we have that
\begin{equation}
\label{eq:lbk_sidm}
    \frac{\rmd L_z}{\rmd t} = 8 \pi^3 \sum_{\mathbf{\ell}} \int \rmd \mathbf{J} \ l_\phi \ \left| \Phi_{1,\mathbf{\ell}} (\mathbf{J},t) \right|^2 \ \mathbf{\ell} \cdot \frac{\partial f_0}{\partial \mathbf{J}} \frac{\nu}{\Delta \Omega^2+ \nu^2} \ ,
\end{equation}
which is equivalent to Equation~(\ref{eq:SIDM_LBK}) apart from minor changes in notation. 

Note that in the collisionless limit $(\nu =0)$, Equation~(\ref{eq:timedep_sidm_lbk}) simplifies to:
\begin{equation}
    \frac{\rmd L_z}{\rmd t} = 8 \pi^3 \sum_{\mathbf{\ell}} \int \rmd \mathbf{J} \ l_\phi \ \left| \Phi_{1,\mathbf{\ell}} (\mathbf{J},t) \right|^2 \ \mathbf{\ell} \cdot \frac{\partial f_0}{\partial \mathbf{J}}  \frac{\sin(\Delta \Omega t)  }{ \Delta \Omega} \ . 
\end{equation}
In the $t \rightarrow \infty$ limit, the above expression further reduces to
\begin{equation}
\label{eq:lbk_cdm}
    \frac{\rmd L_z}{\rmd t} = 8 \pi^4 \sum_{\mathbf{\ell}} \int \rmd \mathbf{J} \ l_\phi \ \left| \Phi_{1,\mathbf{\ell}} (\mathbf{J},t) \right|^2 \ \mathbf{\ell} \cdot \frac{\partial f_0}{\partial \mathbf{J}} \, \delta\left(\Delta \Omega \right) \ ,
\end{equation}
which is equivalent to the classic LBK torque formula of Equation~(\ref{eq:LBK}).

\section{Torque calculation}
\label{ssec:torque_calculation}

We evaluate Equation~(\ref{eq:lbk_sidm}) for a bar-halo system in order to assess the impact of collisions on the torque. The unperturbed halo is assumed to be spherical and isotropic, so that 
\begin{equation}
    \ell \cdot \frac{\partial f_0}{\partial \mathbf{J}} = (\ell \cdot \mathbf{\Omega}) \frac{\partial f_0}{\partial E} \ .
\end{equation}
We focus on the corotation (CR) resonance, which has $\ell=(0,0,2)$, since it makes the largest contribution among all the resonances \citep{Athanassoula2003, Chiba2022}. We make a change of variables from actions $\mathbf{J}=(J_r,J_z,J_\phi)$ to $(E,\kappa,\cos\beta)$ where $E$ is energy, $\kappa=L/L_{\rm c}(E)$ is the total normalized angular momentum, and $\beta$ is the inclination angle with respect to the $x-y$ plane. The Jacobian for this transformation is $\kappa L_{\rm c}(E)^2/\Omega_r$. Therefore we have
that
\begin{equation}
\begin{split}
    \left(\frac{\rmd L_z}{\rmd t} \right)_{\rm CR}= 32 \pi^3 \int \rmd E \int_0^1 \rmd \kappa \int_{0}^1 \rmd (\cos \beta) \frac{\kappa L_{\rm c}(E)^2 \Omega_\phi}{\Omega_r} \\ \times \left| \Phi_{1,\mathbf{\ell}} (\mathbf{J},t) \right|^2   \frac{\partial f_0}{\partial E}  \frac{\nu}{\Delta \Omega^2+ \nu^2} \ ,
\end{split}
\label{eq:lbk_cor_isotropic}
\end{equation}
where we have restricted the integration to prograde orbits ($\cos \beta >0$). 

The computation is carried out using several tools provided by \texttt{Agama}. First, the potentials of the halo and bar are calculated from the $T=12$ Gyr snapshot of the {\tt Fiducial} CDM simulation. We use {\tt Multipole} expansion, restricted to the spherical term only, to construct the halo potential. The spherical approximation simplifies the transformation from $(\boldsymbol{x}, \boldsymbol{v})$ to $(\boldsymbol{J}, \boldsymbol{\theta})$ performed by the {\tt ActionFinder} tool, which constructs an interpolation table for the bijective mapping between $(E,\kappa)$ and $(J_r,J_\phi)$. The corresponding DF of the halo $f_0$ is then constructed numerically via Eddington inversion. The bar potential $\Phi_1(\boldsymbol{x})$ is obtained by constructing a triaxial {\tt Cylspline} potential of the stellar component and subtracting its axisymmetric component. It is then converted to action space as $\Phi_1(\boldsymbol{J}) = \int \rmd\boldsymbol{\theta}\,\Phi\big(\boldsymbol{x}(\boldsymbol{J}, \boldsymbol{\theta})\big)$, where the transformation from $(\boldsymbol{J}, \boldsymbol{\theta})$ to $(\boldsymbol{x}, \boldsymbol{v})$ is performed by the {\tt ActionMapper} tool. In a spherical potential, this amounts to numerically solving the inverse equation for $r(\boldsymbol{\theta})$. 

The orbit-averaged collision frequency $\nu$ for a particle with action $\mathbf{J}$ is approximated as
\begin{equation}
    \nu(\mathbf{J}) \approx  \rho(R_\rmc) \, \sigma_\rmm \, \langle v_{\rm rel} \rangle \ ,
\end{equation}
where $R_\rmc = R_\rmc(E)$ is the radius of a circular orbit with energy $E$, and $\langle v_{\rm rel} \rangle$ is the average relative velocity between two particles, which for a Maxwell--Boltzmann distribution is related to the one-dimensional velocity dispersion $\sigma_{\rm 1D}(r)$ as $\langle v_{\rm rel} \rangle = (4/\sqrt{\pi}) \, \sigma_{\rm 1D}(r)$. 

\begin{figure*}
    \centering
    \includegraphics[width=\textwidth]{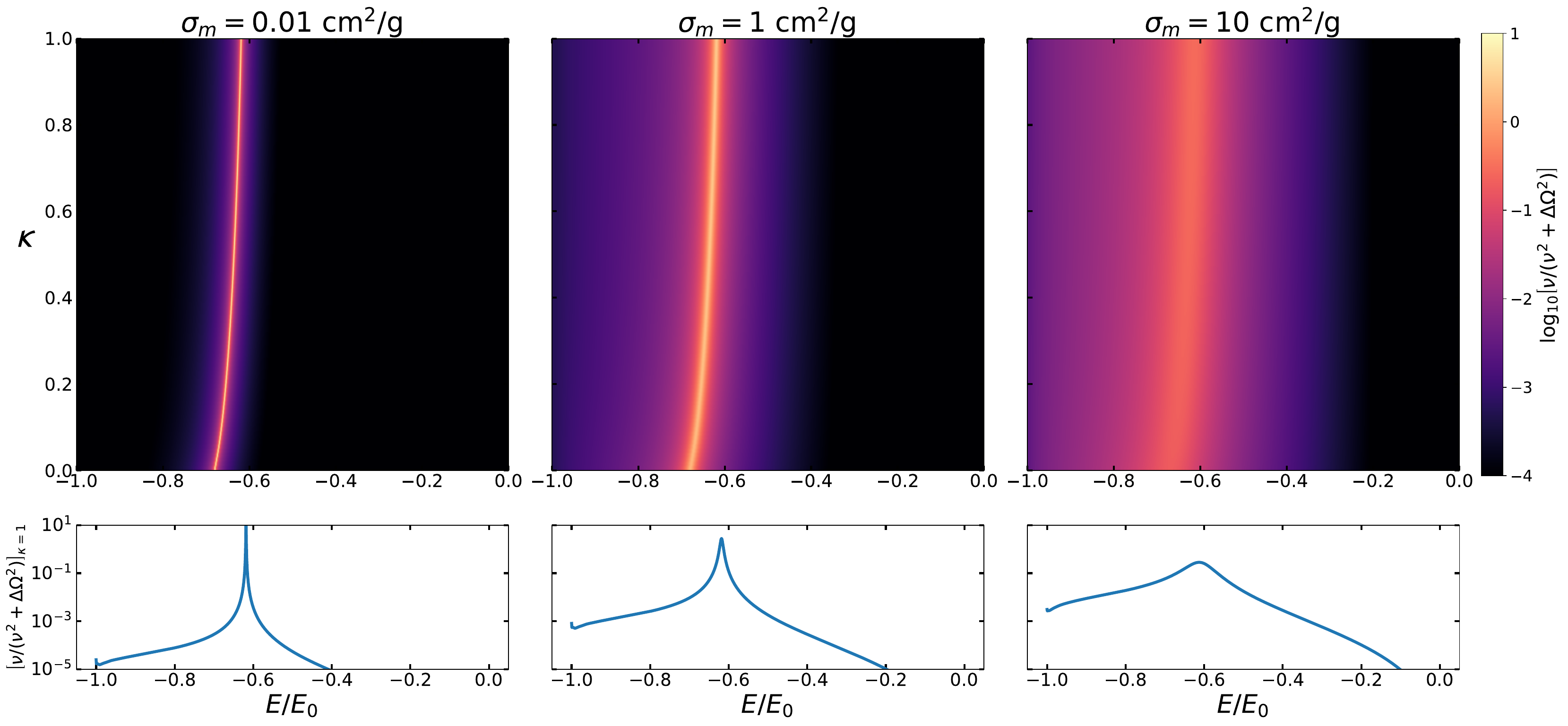}
    \caption{{\it Top panels}: value of the corotation resonance kernel $\nu/(\nu^2 + \Delta \Omega^2)$ in the $E-\kappa$ plane (where $E$ is normalized by $E_0$), for $\sigma_\rmm=0.01 \cmg$ (left), $1 \cmg$ (middle), and $10 \cmg$ (right), for the bar-halo system of the $T=12$ Gyr snapshot of the {\tt Fiducial} CDM simulation. {\it Bottom panels:} one-dimensional cuts of the resonance kernel at $\kappa=1$. For CDM, the kernel would be a Dirac delta function at resonance. For $\sigma_\rmm=0.01 \cmg$, which is essentially CDM-like, the kernel spikes only at the resonance locus where $\Omega_\phi=\Omega_\rmb$. As the cross section is increased, the kernel is broadened, especially across the energy axis. In addition, this broadening is asymmetric with respect to the resonant energy.}
\label{fig:kernel}
\end{figure*}

To demonstrate the resonance broadening, the top panels of Figure~\ref{fig:kernel} show the value of the resonance kernel $\nu/(\nu^2 + \Delta \Omega^2)$ in the $E-\kappa$ plane for $\sigma_\rmm=0.01 \cmg$ (left), $\sigma_\rmm=1\cmg$ (middle), and $\sigma_\rmm=10 \cmg$ (right). The bottom panels show the one-dimensional cut of the kernel at $\kappa=1$ (the curves for other $\kappa$ values are very similar). For $\sigma_\rmm=0.01 \cmg$, which is a very low cross section for which the dynamics are CDM-like, it is clear that the the kernel spikes only at the locus of the corotation resonance ($\Omega_\phi=\Omega_\rmb$). In the $\sigma_\rmm=1\cmg$ case, the kernel has broadened along the $E$ axis, although there is still a clear peak at the resonance. This broadening is even more significant in the $\sigma_\rmm=10 \cmg$ case. Note that, as evident from the bottom panels, the resonance kernel is asymmetric along the $E$ axis, because $\nu$ is larger for particles with lower (more negative) $E$. This asymmetric broadening is key to increasing the torque, since the phase-space regions with lower $E$, which have larger magnitudes of $\partial f_0/\partial E$, will be able to couple to the bar. 

The resonance loci are nearly vertical in the $(E,\kappa)$ plane. Therefore, accurately integrating Equation~(\ref{eq:lbk_cor_isotropic}) requires sufficient resolution along the energy axis. In particular, the difference in $\Omega_\phi$ values between adjacent grid points should be $\lesssim \nu/5$. For the CDM case, in which the resonance kernel is a Dirac delta function, we instead directly evaluate Equation~(\ref{eq:lbk_cdm}) along the corotation resonant locus, defined by $\Delta \Omega = 2\Omega_\phi-2\Omega_\rmb=0$. Since $\Omega_\phi$ is a function only of $E$ and $\kappa$ in a spherical potential, we use Brent's algorithm to calculate $E_{\rm res}(\kappa)$ where $\Delta \Omega=0$. Equation~(\ref{eq:lbk_cdm}) can then be written as
\begin{equation}
\begin{split}    
    \left( \frac{\rmd L_z}{\rmd t}\right)_{\rm CR} = 32 \pi^4 \int_0^1 \rmd \kappa \int_{0}^1 \rmd (\cos \beta) \frac{\kappa L_{\rm c}(E)^2 \Omega_\phi}{\Omega_r} \\ \times \left| \Phi_{1,\mathbf{\ell}} (\mathbf{J},t) \right|^2   \frac{\partial f_0}{\partial E}  \left| \frac{\partial (\Delta \Omega)}{\partial E} \right|^{-1} \ ,
\end{split}
\end{equation}
where the integrand is evaluated at $E_{\rm res}(\kappa)$. 

\section{Pendulum approximation and fast-slow limits}
\label{ssec:pendulum}

As mentioned in Section~\ref{ssec:theory}, the LBK torque (both the collisionless and collisional variants) is only valid in the fast regime, in which the patterns speed of the bar is changing fast enough such that no orbit remains near the separatrix of a resonance long enough for non-linear effects to build up. \citet{TW84} refer to this process as ``sweeping across the resonances". In the opposite ``slow regime'', particles can get trapped into librating orbits, which continuously exchange angular momentum with the bar \citep{TW84, Banik2022, Chiba2022}. In the long-term limit, phase-mixing among such librating orbits will drive their net torque contribution to zero \citep{Chiba2023,Hamilton2024, Kodama2026}. This corresponds to a flattening of the system's DF in the vicinity of the resonance. However, \citet{Hamilton2023} showed that collisionality will suppress phase-mixing, maintaining a non-zero DF gradient across the resonance and therefore a non-vanishing torque \citep[see also][]{vdb2025, Kodama2026}. Hence in the slow regime, collisions prevent, or at least delay, the saturation of the growing bar mode.  

In order to verify that the bars in our simulations are in the fast limit, we use the pendulum approximation to calculate the rate of separatrix evolution \citep{Neishtadt1975, Henrard1982}. In the vicinity of a resonance, defined by $\mathbf{\ell} \cdot \mathbf{\Omega} - l_\phi \Omega_\rmb=0$, we apply a canonical transformation to a new set of slow and fast action-angle variables:
\begin{equation}
\begin{split}
    \theta_\rms = \mathbf{\ell} \cdot \mathbf{\Omega} - l_{\phi} \Omega_\rmb t\,, \quad \theta_{\rmf_1} = \theta_r\,, \quad \theta_{\rmf2}=\theta_z \ , \\
    J_\rms = J_\phi\,, \quad J_{\rmf1} = J_r - \frac{l_r}{l_\phi} J_\phi \,, \quad J_{\rmf2} = J_z - \frac{l_z}{l_\phi} J_\phi \ , 
\end{split}    
\end{equation}
where the subscripts `s' and `f' stand for `slow' and `fast' respectively.

The fast angles evolve on an orbital timescale, whereas the slow angle evolves over the much longer timescale $\sim (\mathbf{\ell} \cdot \mathbf{\Omega} - l_\phi \Omega_\rmb)^{-1}$. Therefore, the Hamiltonian can be averaged over the fast angles, and correspondingly the fast actions are conserved. In slow action-angle space, the Hamiltonian can be Taylor expanded about the resonant slow action $(J_{\rms,{\rm res}})$ to obtain
\begin{equation}
    H = \frac{1}{2} G \left( J_\rms - J_{\rms,{\rm res}} \right) - F \cos \left( \theta_\rms - \theta_{\rms,{\rm res}} \right) \ ,
\end{equation}
\citep[e.g.,][]{TW84, Chiba2022}. This is the Hamiltonian for a one-dimensional pendulum with coefficients
\begin{equation}
    G = \frac{\partial^2 H}{\partial J_\rms^2} = \frac{\partial (\mathbf{\ell} \cdot \mathbf{\Omega})}{\partial J_\rms} \ ,  
\end{equation}
and
\begin{equation}
    F=-2 \left| \Phi_{1,(1,0,0)} \right| \ , 
\end{equation}
where $\Phi_{1,\ell}$ is the Fourier transform of the bar potential in $(\theta_\rms, \theta_{\rmf1}$, $\theta_{\rmf2})$ coordinates.

The libration frequency at resonance is 
\begin{equation}
    \Omega_{\rm lib} = \sqrt{GF} \ , 
\end{equation}
and the total area enclosed by the separatrix is
\begin{equation}
    S=16 \sqrt{\frac{\left| F \right|}{G}} \ .
\end{equation}

The rate of separatrix evolution can be characterized by the dimensionless parameter
\begin{equation}
    \Gamma=\frac{\left| \dot{S}\right|}{\Omega_{\rm lib}\,S} \ ,
\end{equation}
where $\Gamma>1$ and $\Gamma<1$ correspond to the fast and slow regimes, respectively. 

We calculate the pendulum coefficients in our simulations for the corotation resonance. The frequencies $\mathbf{\Omega}$ are calculated as described in Section~\ref{ssec:phase_space}, and the bar potential and its Fourier transform are calculated as described in Appendix~\ref{ssec:torque_calculation}.

We find that the value of $\Gamma$ is $>1$ at all times in the {\tt Fiducial} suite of simulations, with typical values around $\Gamma \sim 3$. Therefore, we conclude that the bars (at least in the {\tt Fiducial} runs) are in the fast regime. This implies that the accelerated bar growth in SIDM is primarily due to the resonance broadening effect and not due to suppression of phase-mixing (although we suspect that the latter still contributes to some extent). 


\bsp	
\label{lastpage}
\end{document}